\documentclass[english]{IEEEtran}
\usepackage[T1]{fontenc}
\usepackage[latin9]{inputenc}
\usepackage{mathrsfs}
\usepackage{amsthm}
\usepackage{amsmath}
\usepackage{amssymb}
\usepackage{graphicx}
\usepackage{esint}

\makeatletter

\providecommand{\tabularnewline}{\\}

  \theoremstyle{definition}
  \newtheorem{example}{\protect\examplename}
  \theoremstyle{remark}
  \newtheorem{rem}{\protect\remarkname}
  \theoremstyle{definition}
  \newtheorem{defn}{\protect\definitionname}
  \theoremstyle{plain}
  \newtheorem{thm}{\protect\theoremname}
  \theoremstyle{plain}
  \newtheorem{cor}{\protect\corollaryname}
  \theoremstyle{plain}
  \newtheorem{lem}{\protect\lemmaname}
  \theoremstyle{plain}
  \newtheorem{prop}{\protect\propositionname}

\newtheorem{assumption}{Assumption}
\usepackage{subfig}

\makeatother

\usepackage{babel}
\providecommand{\corollaryname}{Corollary}
\providecommand{\definitionname}{Definition}
\providecommand{\examplename}{Example}
\providecommand{\lemmaname}{Lemma}
\providecommand{\propositionname}{Proposition}
\providecommand{\remarkname}{Remark}
\providecommand{\theoremname}{Theorem}

\begin{document}

\title{Cache-Enabled Opportunistic Cooperative MIMO for Video Streaming
in Wireless Systems }

\author{{\normalsize{An Liu, }}\textit{\normalsize{Member IEEE}}{\normalsize{,
and Vincent Lau,}}\textit{\normalsize{ Fellow IEEE}}{\normalsize{,\\Department
of Electronic and Computer Engineering, Hong Kong University of Science
and Technology}}}
\maketitle
\begin{abstract}
We propose a \textit{cache-enabled opportunistic} cooperative MIMO
(CoMP) framework for wireless video streaming. By caching a portion
of the video files at the relays (RS) using a novel \textit{MDS-coded
random cache} scheme, the base station (BS) and RSs opportunistically
employ CoMP to achieve spatial multiplexing gain without expensive
payload backhaul. We study a two timescale joint optimization of power
and cache control to support real-time video streaming. The cache
control is to create more CoMP opportunities and is adaptive to the
long-term popularity of the video files. The power control is to guarantee
the QoS requirements and is adaptive to the channel state information
(CSI), the \textit{cache state} at the RS and the queue state information
(QSI) at the users. The joint problem is decomposed into an \textit{inner
power control problem} and an \textit{outer cache control problem}.
We first derive a closed-form power control policy from an \textit{approximated
Bellman equation}. Based on this, we transform the outer problem into
a convex stochastic optimization problem and propose a stochastic
subgradient algorithm to solve it. Finally, the proposed solution
is shown to be asymptotically optimal for high SNR and small timeslot
duration. Its superior performance over various baselines is verified
by simulations.\end{abstract}
\begin{IEEEkeywords}
Wireless video streaming, Dynamic cache control, Opportunistic CoMP,
Power control

\thispagestyle{empty}
\end{IEEEkeywords}

\section{Introduction}

Nowadays, more and more high data rate applications, such as video
streaming, are supported by wireless networks. It is envisioned that
future 5G wireless systems should offer 1000X increase in capacity.
One key technology to achieve this challenging goal is by means of
cooperative MIMO (CoMP) \cite{somekh2009cooperative,Irmer_Comm11_CoMPsurvey}
because it can transform the cross-link interference into spatial
multiplexing gain. However, CoMP technique requires high capacity
backhaul for payload exchange between BSs, which is a cost bottleneck
for large scale deployment of CoMP in dense small cell networks. On
the other hand, cooperative relay has been proposed as an alternative
technique to improve the performance of wireless networks without
expensive backhaul. However, it can only contribute to coverage and
SNR gain and the overall capacity gain is marginal \cite{Zhang_junsheng_IT05_power_allocation_relay,Zhang_IT05_Capacity_of_MIMO_relay_ch}.

In this paper, we propose a novel solution framework, namely the \textit{cache-enabled
opportunistic CoMP}, to fully unleash the benefits of CoMP without
expensive payload backhaul. By introducing cache at the relay (RS),
we can opportunistically transform a relay channel into a CoMP broadcast
channel and achieve not only SNR gain but also spatial multiplexing
gain. Due to the absence of payload backhaul, such a solution is very
attractive for dense small cell networks. In traditional wireless
networks, the information bits are treated as random bits and hence,
the capacity is limited by the underlying network topology. However,
in future wireless networks, the source should be viewed as \textit{content}
rather than random bits. For example, it is envisioned that a significant
portion of the 1000X capacity demand in 5G systems comes from video
streaming applications. As such, a large portion of video traffic
corresponds to a few popular video files which are requested by many
users. There are huge correlations between the information accessed
by users and such information is \textit{cachable} at the RS. If a
user accesses a content which exists in the RS cache, the BS and the
RS can engage in CoMP and therefore, enjoy spatial multiplexing gain
in addition to SNR gain. Hence, with proper caching strategy, a large
portion of backhaul usage can be replaced by the cache. Since the
cost of hard disks is much lower than the cost of optical fiber backhaul,
the proposed solution is very cost effective. 

Yet, the opportunity of CoMP in the information flows depends heavily
on the dynamic caching strategy. Caching has been widely used in fixed
line P2P systems \cite{Hefeeda_TACM08_P2P,Kozat_TOM09_P2P} and content
distribution networks (CDNs) \cite{Shen_TOM04_CDN,Borst_INFOCOM10_CDN}.
The benefits of caching in these cases mainly come from reducing the
number of hops from the content source to the consumers. However,
we have a fundamentally different caching cost and reward dynamics
in the wireless scenario. In addition to the common advantage of bringing
the content close to the consumers, there is a unique advantage of
cache-induced topology change in the physical layer (dynamic CoMP
opportunity) of the wireless systems. As such, one cannot directly
adopt conventional caching algorithms in the fixed line networks.
In \cite{Caire_INFOCOM12_femtocache}, a FemtoCaching scheme is proposed
by using small BSs called helpers with high storage capacity cache
popular video files. However, the FemtoCaching scheme does not consider
cache-enabled opportunistic CoMP among the BS and local helpers. Hence,
the cache control in \cite{Caire_INFOCOM12_femtocache} is independent
of the physical layer and is fundamentally different from our case
where the cache control and physical layer are coupled together. In
our recent paper \cite{Liu_TSP13_CacheIFN}, we studied a mixed-timescale
optimization of MIMO precoding and cache control for MIMO interference
networks employing the cache-enabled opportunistic CoMP. However,
the cache control and MIMO precoding algorithms in \cite{Liu_TSP13_CacheIFN}
can only be used to guarantee the physical layer QoS (fixed data rate
requirement for each user) and the MIMO precoder is not adaptive to
the QSI at the mobile users.

In this paper, we study two timescale joint optimization of power
and cache control to support real-time video-on-demand (VoD) applications.
The role of cache control is to create (induce) more CoMP opportunities
and is adaptive to long-term popularity of the video files (long-term
control). The role of dynamic power control is to exploit the CoMP
opportunities (induced by the cache) to strike a balance between capacity
gain and the \textit{urgency} between information flows. As such,
it is adaptive to the instantaneous CSI, the \textit{cache state}
at the RS and the QSI at the mobile users. There are several technical
challenges to be addressed.
\begin{itemize}
\item \textbf{Cache Size Requirement}: The performance gain of the proposed
scheme depends heavily on the CoMP opportunity, which in turn depends
on the cache size and cache algorithm. The RS usually does not have
enough cache to store all the video files. As will be shown in Example
\ref{Naive-cahce-scheme}, when brute force caching is used, even
if a significant portion of the video files are cached at RS, the
CoMP opportunity can still be very small and this is highly undesirable.
\item \textbf{Unknown Video File Popularity}: In many cases, the RS does
not have the knowledge of the\textbf{ }popularity of the video files.
Hence, the cache control must have the ability to automatically learn
the popularity of video files.
\item \textbf{Challenge of QoS Requirement}: It is important to dynamically
control the transmit powers at the BS and the RS based on the CSI,
cache state (reveals good opportunities of transmission) and the QSI
(reveals urgency of the data flow) at the mobile in order to maintain
low interruption probability during video playback. Yet, the optimization
belongs to the infinite horizon Markov Decision Process (MDP) \cite{Bertsekas_MIT07_DynProg,Cao_Springer08_SLO},
which is well-known to be very challenging.
\item \textbf{Complex Coupling between Cache and Power Control}: The cache
control will affect the physical layer dynamics seen by power control
due to different CoMP opportunities. Furthermore, the optimization
objective of the cache control (long-term) depends on the optimal
power solution as well as the average cost of the MDP. However, there
are no closed-form characterizations for them.
\end{itemize}

To address the above challenges, we first propose a novel cache data
structure called \textit{MDS-coded random cache} which can significantly
improve the probability of CoMP. We then exploit the timescale separations
of the control variables to decompose the problem into an \textit{inner
power control problem} and an \textit{outer cache control problem}.
To solve the inner power control problem, we exploit a specific problem
structure and derive an approximate Bellman equation. Based on this,
we derive a closed-form approximation for the value function, the
power control policy as well as the average cost. To solve the outer
cache control problem, we exploit these closed form approximations
for the inner problem and transform the outer problem into a convex
stochastic optimization problem. Then we propose a stochastic subgradient
algorithm to solve it. We show that the proposed solution is asymptotically
optimal for high SNR and small timeslot duration. Finally, we illustrate
with simulations that the proposed solution achieves significant gain
over various baselines.

\textit{Notation}\emph{s}: The superscript $\left(\cdot\right)^{\dagger}$
denotes Hermitian. The notation $1\left(\cdot\right)$ denote the
indication function such that $1\left(E\right)=1$ if the event $E$
is true and $1\left(E\right)=0$ otherwise. The notation $\textrm{span}\left(\mathbf{A}\right)$
represents the subspace spanned by the columns of a matrix $\mathbf{A}$.

\section{System Model\label{sec:System-Model}}

In this section, we introduce the architecture of the video streaming
system, the physical layer (opportunistic CoMP), the MDS-coded random
cache scheme that supports opportunistic CoMP, and the queue model.

\subsection{Architecture of Video Streaming System}

\begin{figure}
\begin{centering}
\textsf{\includegraphics[clip,width=90mm]{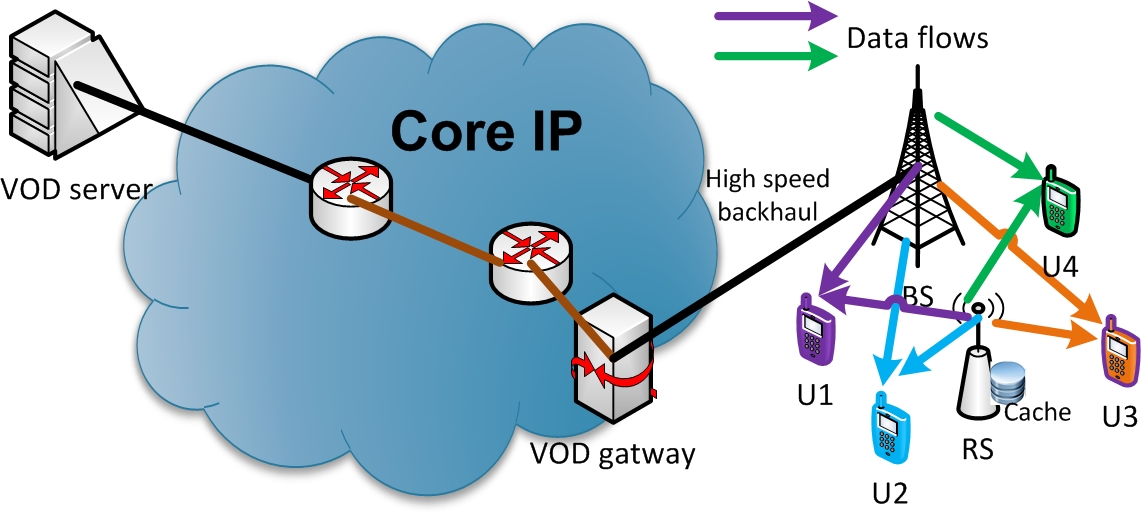}}
\par\end{centering}

\caption{\label{fig:system_model}The architecture of the video streaming system. }
\end{figure}

The architecture of the video streaming system is illustrated in Fig.
\ref{fig:system_model}. There are $L$ video files on the VoD server.
The size of the $l$-th video file is $F_{l}$ bits. For simplicity,
we assume that the $L$ video files have the same streaming rate denoted
by $\mu_{0}$ (bits/s). There are $2M$ users streaming video files
from the VoD server via a radio access network (RAN) consisting of
an $M$-antenna BS and an $M$-antenna RS%
\footnote{For clarity, we consider the case with one RS and $2M$ users. However,
the solution framework can be easily extended to the case with multiple
RSs and more than $2M$ users.%
}. The index of the video file requested by the $k$-th user is denoted
by $\pi_{k}$. Define $\pi=\left\{ \pi_{1},...,\pi_{2M}\right\} $
as the user request profile (URP). We have the following assumption
on URP.

\begin{assumption}[URP Assumption]\label{asm:URP}The URP $\pi\left(t\right)$
is a slow ergodic random process (i.e., $\pi\left(t\right)$ remains
constant for a large number of time slots) according to a general
distribution.

\end{assumption}

Video packets are delivered to the BS from the VoD gateway via the
high speed backhaul as illustrated in Fig. \ref{fig:system_model}.
On the other hand, the RS has no fixed line backhaul connection with
the BS and it is equipped with a cache. In this paper, the time is
partitioned into time slots indexed by $t$ with slot duration $\tau$.

\subsection{Cache-enabled Opportunistic CoMP\label{sub:Cache-enabled-Opportunistic-CoMP}}

The RAN is the performance bottleneck of the system. Without cache
at the RS, the RAN forms a MIMO relay channel and the conventional
relay techniques can only contribute to coverage and SNR gain. In
this section, we propose a cache-enabled opportunistic CoMP which
can opportunistically use the cached video packets at the RS to transform
a relay channel into a CoMP broadcast channel as illustrated in Fig.
1. As a result, the RAN can enjoy the additional spatial multiplexing
gain \cite{Zheng_IT02_Multiplexing_tradeoff} without expensive backhauls.

We first define some system states. Let $\mathbf{h}_{k}\in\mathbb{C}^{2M}$
denote the channel vector between user $k$ and the $2M$ transmit
antennas at the BS and RS. Let $\overline{\mathbf{h}}_{k}\in\mathbb{C}^{M}$
denote the sub channel vector between user $k$ and the $M$ transmit
antennas at the BS. Let $\mathbf{H}=\left\{ \mathbf{h}_{k},\forall k\right\} $
denote the global CSI. We have the following assumption on the CSI
$\mathbf{H}$. 

\begin{assumption}[Channel model]\label{asm:forY}$\mathbf{h}_{k}\left(t\right)$
remains constant within a time slot but is i.i.d. w.r.t. time slots
$t$ and user index $k$. Specifically, $\mathbf{h}_{k}\left(t\right)$
has i.i.d. complex Gaussian entries of zero mean and unit variance.

\end{assumption}

The impact of caching at RS on the physical layer is summarized by
the \textit{cache state} defined as $S\in\left\{ 0,1\right\} $, where
$S=1$ means that the current payload data requested by the $2M$
users is in the RS cache and thus it is possible for the BS and RS
to cooperatively transmit the payload data to the users, and $S=0$
means that the users can only be served by the BS.

Then we elaborate the proposed cache-enabled opportunistic CoMP. There
are two transmission modes depending on the cache state S.

\textbf{Mode 0}: If $S=0$, $M$ users are selected randomly for transmission
from the $2M$ users. For convenience, let $\Theta$ denote the set
of selected users. Then for any $k\in\Theta$, the BS transmits a
signal vector $\overline{\mathbf{v}}_{k}\sqrt{p_{k}}s_{k}\in\mathbb{C}^{M}$
using the $M$ transmit antennas at the BS, where $p_{k}$ is the
transmit power for user $k$; $s_{k}$ is the data symbol for user
$k$; and $\overline{\mathbf{v}}_{k}\in\mathbb{C}^{M}$ with $\left\Vert \overline{\mathbf{v}}_{k}\right\Vert =1$
is the zero-forcing (ZF) beamforming vector which is obtained by the
projection of $\overline{\mathbf{h}}_{k}$ on the orthogonal complement
of the subspace $\textrm{span}\left(\left[\overline{\mathbf{h}}_{k^{'}}\right]_{k^{'}\in\Theta\backslash\left\{ k\right\} }\right)$.

\textbf{Mode 1}: If $S=1$, all the $2M$ users are selected and served
using CoMP between the BS and the RS. For any $k$, the BS and RS
jointly transmit a signal vector $\mathbf{v}_{k}\sqrt{p_{k}}s_{k}\in\mathbb{C}^{2M}$
using the $2M$ transmit antennas at both BS and RS, where $\mathbf{v}_{k}\in\mathbb{C}^{2M}$
with $\left\Vert \mathbf{v}_{k}\right\Vert =1$ is the ZF beamforming
vector and it is obtained by the projection of $\mathbf{h}_{k}$ on
the orthogonal complement of the subspace $\textrm{span}\left(\left[\mathbf{h}_{k^{'}}\right]_{k^{'}\in\Theta\backslash\left\{ k\right\} }\right)$.

\begin{figure}
\begin{centering}
\textsf{\includegraphics[clip,width=90mm]{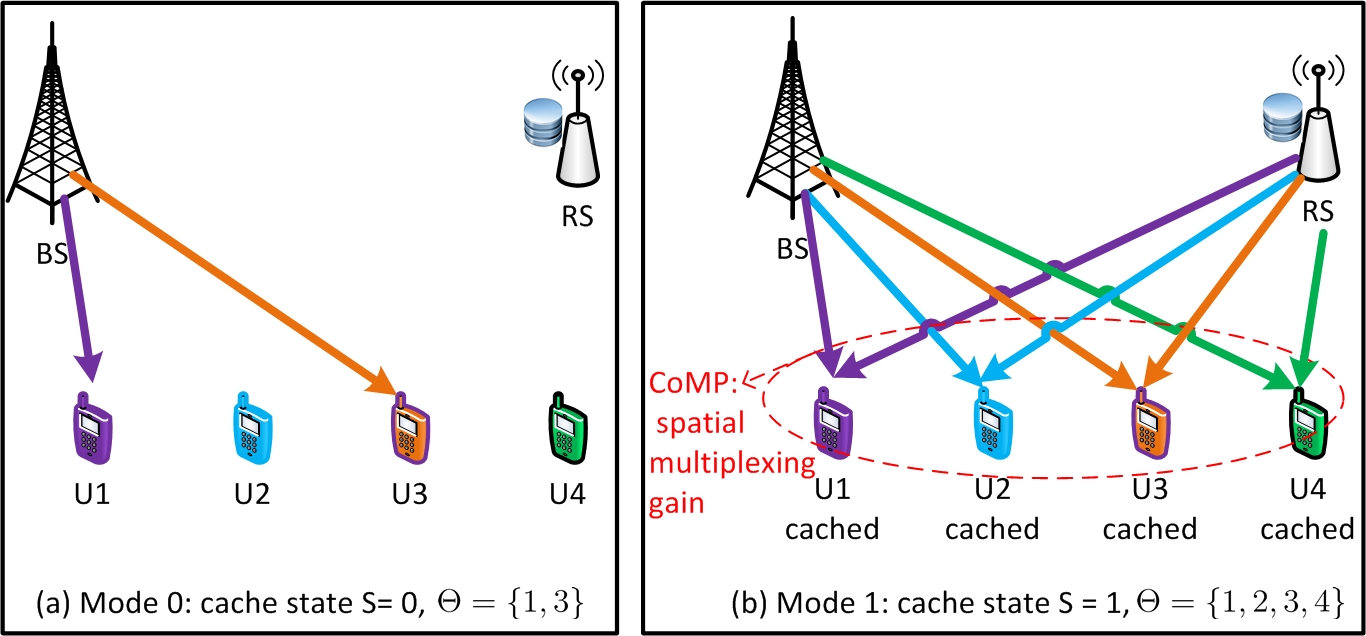}}
\par\end{centering}

\caption{\label{fig:CocaMIMO}Illustration of cache-enabled opportunistic CoMP
for a RAN with $M=2$. }
\end{figure}

Fig. \ref{fig:CocaMIMO} illustrates two examples of the data flows
under the proposed cache-enabled opportunistic CoMP with different
cache states $S$ (or transmission modes). Note that using ZF beamforming,
there is no interference among different data flows. Define the \textit{effective
channel} $\widetilde{g}_{k}$ of user $k$ as
\[
\widetilde{g}_{k}=\begin{cases}
\mathbf{h}_{k}^{\dagger}\mathbf{v}_{k}, & \textrm{if}\: S=1,\\
\overline{\mathbf{h}}_{k}^{\dagger}\overline{\mathbf{v}}_{k}, & \textrm{if}\: S=0,\:\textrm{and}\: k\in\Theta,\\
0, & \textrm{otherwise}.
\end{cases}
\]
Then the received signals at user $k$ can be expressed as
\begin{eqnarray*}
y_{k} & = & \widetilde{g}_{k}\sqrt{p_{k}}s_{k}+z_{k},
\end{eqnarray*}
where $z_{k}\sim\mathcal{CN}\left(0,1\right)$ is the AWGN noise.
Note that the effective channels $\left\{ \widetilde{g}_{k},\forall k\right\} $
are determined by the global CSI $\mathbf{H}$ and the cache state
$S$. Let $g_{k}=\left|\widetilde{g}_{k}\right|^{2}$. Then for given
CSI $\mathbf{H}$ and cache state $S$, the data rate of user $k$
is given by
\[
r_{k}\left(g_{k},p_{k}\right)=B_{W}\textrm{log}_{2}\left(1+g_{k}p_{k}\right),
\]
where $B_{W}$ is the bandwidth of the system.

As illustrated in Fig. \ref{fig:CocaMIMO}, when $S=1$, there is
spatial multiplexing gain due to cache-enabled CoMP transmission.
The overall performance gain depends heavily on the probability of
$S=1$. In the next section, we propose a novel MDS-coded random cache
scheme which makes best use of the RS cache to increase $\textrm{Pr}\left[S=1\right]$.

\subsection{MDS-coded Random Cache\label{sub:Outline-of-cache}}

We use an example to show that, with a naive cache scheme, the CoMP
opportunity ($S=1$) can be very small even if a significant portion
of the video files are stored at the RS cache.
\begin{example}
[A Naive Cache Scheme]\label{Naive-cahce-scheme}Suppose that there
are $L=2M=8$ video files with equal size of $F$ bits and the $k$-th
video file is requested by the $k$-th user. The RS randomly stores
half of the video packets (i.e., $0.5F$ bits) for each video file.
The probability that the packets requested by a single user are in
the RS cache is approximately 0.5. However, the probability of $S=1$
(i.e., the packets requested by all the $2M$ users are in the RS
cache) is only $0.5^{2M}=0.0039$.
\end{example}

Hence, a more intelligent cache scheme is needed. In the following,
we propose a novel MDS-coded random cache scheme which can significantly
improve the probability of CoMP.

\begin{figure}
\begin{centering}
\textsf{\includegraphics[clip,width=90mm]{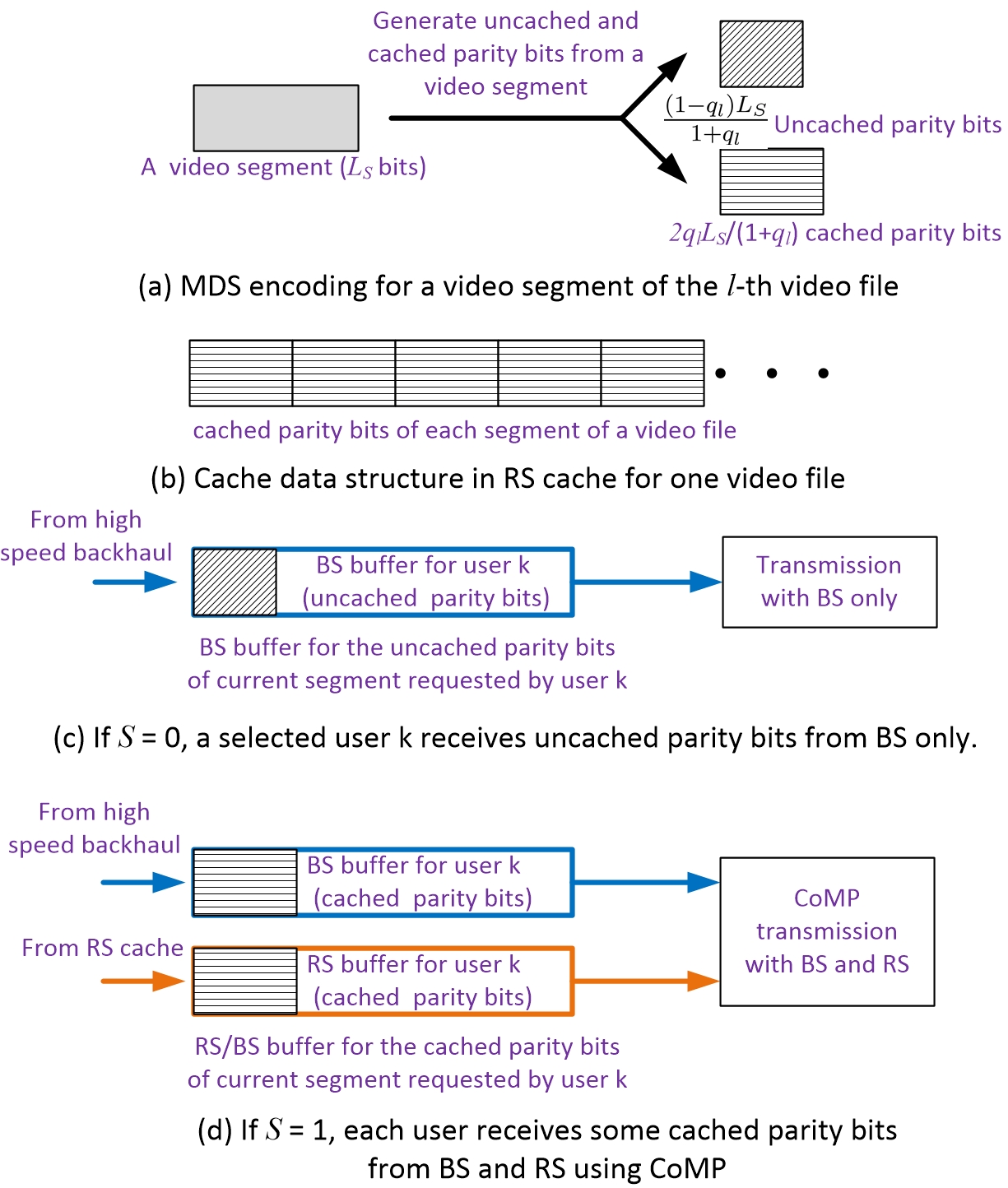}}
\par\end{centering}

\caption{\label{fig:MDS-cache}MDS-coded cache data structure and cache usage.}
\end{figure}

\subsubsection*{Cache Data Structure}

Each video file is divided into segments. Each segment contains $L_{S}\gg\mu_{0}\tau$
bits and it is encoded into $L_{S}$ parity bits using an ideal MDS
rateless code as illustrated in Fig. \ref{fig:MDS-cache}-(a), where
$q_{l}\in\left[0,1\right]$ is called the \textit{cache control variable}.
An MDS rateless code generates an arbitrarily long sequence of parity
bits from an information packet of $L_{S}$ bits, such that if the
decoder obtains any $L_{S}$ parity bits, it can recover the original
$L_{S}$ information bits. In practice, the MDS rateless code can
be implemented using Raptor codes \cite{Shokrollahi_TIT06_Raptorcode}
at the cost of a small redundancy overhead. The RS stores $\frac{2q_{l}L_{S}}{1+q_{l}}$
parity bits for every segment of the $l$-th video file as illustrated
in Fig. \ref{fig:MDS-cache}-(b). Such a cache data structure is more
flexible than the naive cache scheme in Example \ref{Naive-cahce-scheme}
in the sense that we can control when to use the cached data.

\subsubsection*{Random Cache Usage}

The cache usage is determined by the cache state $S$ as illustrated
in Fig. \ref{fig:MDS-cache}-(c,d). If $S=1$, the BS and RS employ
CoMP to jointly transmit the cached parity bits%
\footnote{We assume that the BS has a copy of the cached parity bits.%
} to the $2M$ users using the Mode 1 ZF scheme described in Section
\ref{sub:Cache-enabled-Opportunistic-CoMP}. Otherwise, only $M$
users can be served by the BS using the Mode 0 ZF scheme. Using the
above MDS-coded cache data structure, we can actively control the
cache state $S$ for each time slot. The key to increasing the probability
of CoMP in the system is to align the transmissions of the cached
data for different users as much as possible. Specifically, for given
cache control vector $\mathbf{q}=\left[q_{1},...,q_{L}\right]$ and
$\pi$, let $q_{\textrm{min}}=\underset{1\leq k\leq2M}{\textrm{min}}\left\{ q_{\pi_{k}}\right\} $.
Then the cache state $S$ at each time slot is independently generated
according to a Bernoulli distribution with $\textrm{Pr}\left[S=1\right]=q_{\textrm{min}}$.
\begin{rem}
[Cache Underflow]If $S=1$ and there is no untransmitted parity
bits left in the RS cache for the segment requested by user $k$,
a \textit{cache underflow} event occurs for user $k$. Later in Proposition
\ref{prop:cacheunderflow}, we will show that the cache underflow
probability tends to zero as the segment size $L_{S}\rightarrow\infty$.
Hence, the effect of cache underflow is negligible for large $L_{S}$. 
\end{rem}

The following example illustrates the advantage of MDS-coded random
cache scheme.
\begin{example}
[Advantage of MDS-coded random cache]\label{Advantage-of-MDS-coded}Consider
the setup in Example \ref{Naive-cahce-scheme}. We have $q_{\pi_{k}}=0.5,\forall k$
and thus the probability of $S=1$ is $0.5$, which is much larger
than that of the naive scheme in Example 1 ($0.0039$).
\end{example}

Compared with the naive caching scheme in Example \ref{Naive-cahce-scheme},
the probability of CoMP transmission ($S=1$) under the proposed MDS-coded
random cache is $\textrm{min}_{k}\left\{ q_{\pi_{k}}\right\} $ versus
$\prod_{k=1}^{2M}q_{\pi_{k}}$. This represents a first order improvement
in the opportunity of CoMP gain. Yet, there is a fundamental tradeoff
between the performance gain and the RS cache size. Intuitively, the
more popular the video file is, the larger portion of its parity bits
should be stored in the RS cache to increase the CoMP probability.
Hence, the value of $\mathbf{q}$ must be carefully controlled to
achieve the best tradeoff among performance and required RS cache
size. As such, the cache control variable is parameterized by the
vector $\mathbf{q}=\left[q_{1},...,q_{L}\right]$.

\subsection{Queue Dynamics and QoS Metric}

\begin{figure}
\begin{centering}
\textsf{\includegraphics[clip,width=90mm]{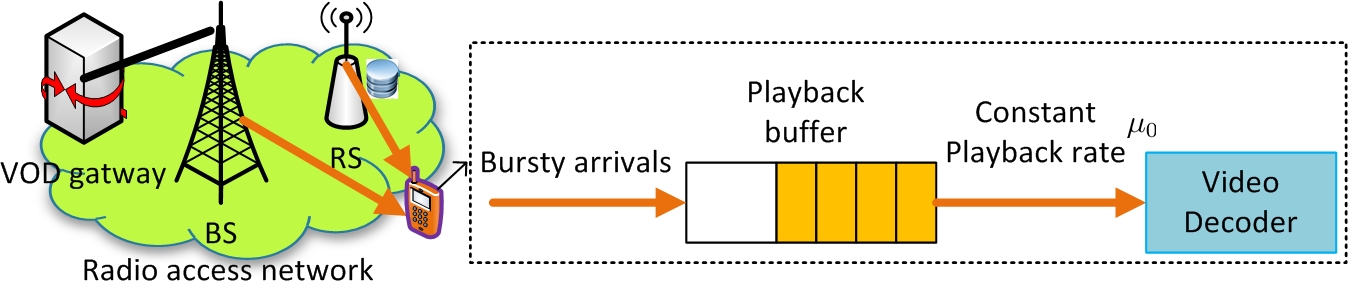}}
\par\end{centering}

\caption{\label{fig:Playbuffer}Playback buffer for video streaming at each
user. }
\end{figure}

Each user maintains a queue for video playback as illustrated in Fig.
\ref{fig:Playbuffer}. Let $Q_{k}\left(t\right)\in\mathcal{Q}$ denote
the QSI (number of bits) of the playback buffer at the $k$-th user,
where $\mathcal{Q}=\left[0,\infty\right)$ is the QSI state space.
Let $\mathbf{Q}(t)=\left(Q_{1}\left(t\right),...,Q_{K}\left(t\right)\right)\in\boldsymbol{\mathcal{Q}}\triangleq\mathcal{Q}^{2M}$
denote the global QSI. The playback rate (departure rate) of user
$k$ depends on the amount of video data in the playback buffer $Q_{k}(t)$
and is given by:
\begin{equation}
\mu_{k}\left(Q_{k}\left(t\right)\right)=\begin{cases}
\frac{Q_{k}\left(t\right)\mu_{0}}{W_{L}}, & Q_{k}\left(t\right)<W_{L},\\
\mu_{0}, & Q_{k}\left(t\right)\geq W_{L},
\end{cases}\label{eq:playrate}
\end{equation}
where $W_{L}>\mu_{0}\tau$ is a parameter. If $Q_{k}\left(t\right)\geq W_{L}$,
a constant playback rate $\mu_{0}$ (which is equal to the streaming
rate) can be maintained. Otherwise, the playback rate is $\mu_{k}=\frac{Q_{k}\left(t\right)\mu_{0}}{W_{L}}$
to avoid buffer underflow. The dynamics of the playback buffer at
user $k$ is given by
\begin{equation}
Q_{k}\left(t+1\right)=Q_{k}\left(t\right)+\left(r_{k}\left(g_{k}\left(t\right),p_{k}\left(t\right)\right)-\mu_{k}\left(Q_{k}\left(t\right)\right)\right)\tau.\label{eq:Qdyn}
\end{equation}
Note that the playback rate in (\ref{eq:playrate}) ensures that $Q_{k}\left(t\right)\geq0$.

Since the video playback quality of a user will be degraded whenever
$Q_{k}<W_{L}$, we define the QoS requirement of each user in terms
of the \textit{interruption probability:}
\begin{equation}
I_{k}=\textrm{Pr}\left[Q_{k}<W_{L}\right].\label{eq:interProb}
\end{equation}
We will elaborate more about the interruption cost together with other
system costs in Section \ref{sub:Performance-Costs}.

\section{Optimization Problem Formulation\label{sec:Optimization-Problem-Formulation}}

\begin{figure}
\begin{centering}
\includegraphics[width=90mm]{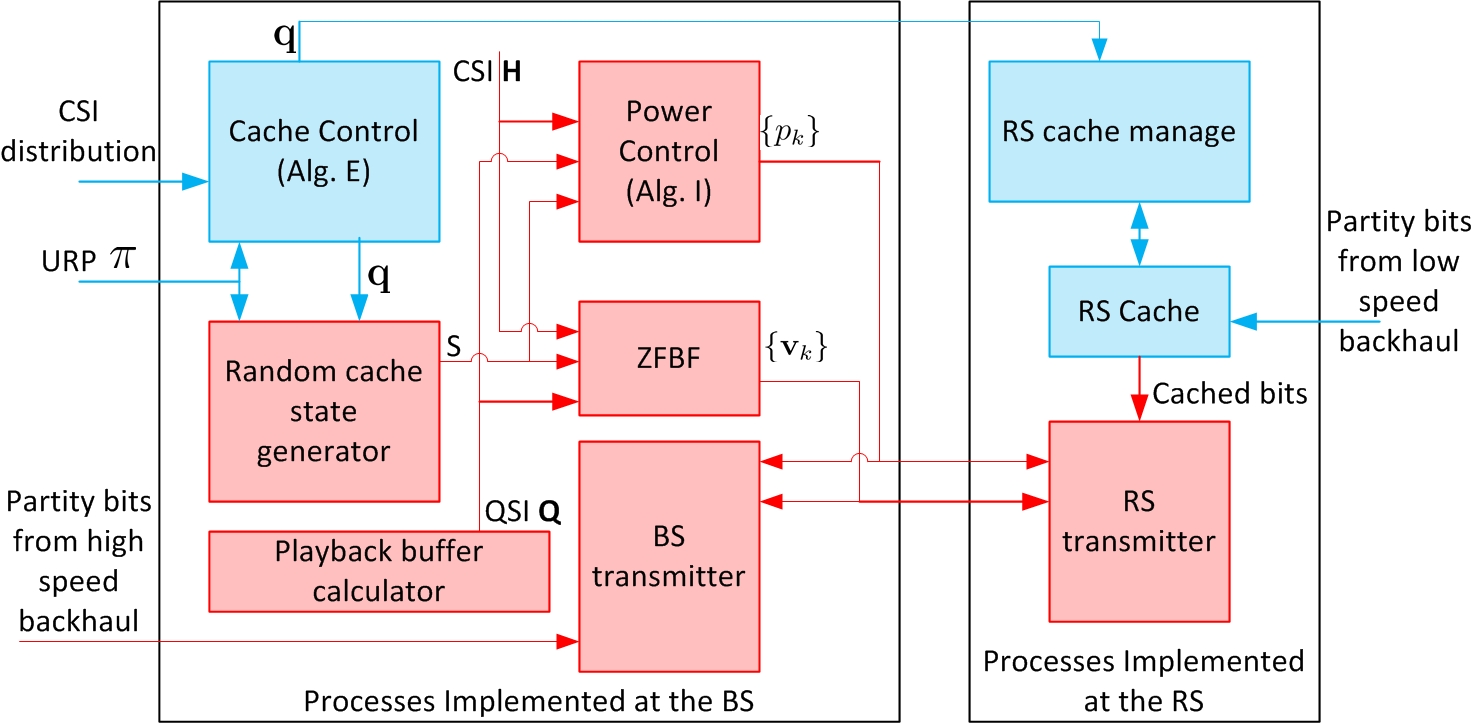}
\par\end{centering}

\caption{\label{fig:moduleconn}{\small{Summary of overall solution and the
inter-relationship of the algorithm components. The blue / red blocks
represent long timescale / short timescale processes. The blue / red
arrows represent long-term / short-term signaling.}}}
\end{figure}

In this section, we introduce the power control policy and formulate
a two timescale optimization problem for video streaming under cache-enabled
opportunistic CoMP. The control variables are partitioned into \textit{long-term}
and \textit{short-term} control variables. As illustrated in Fig.
\ref{fig:moduleconn}, the long-term control variables (cache control
variables $\mathbf{q}$) are adaptive to the distribution of the URP
$\pi$ to induce CoMP opportunity. The short-term control variables
(power control variables $\left\{ p_{k},\forall k\right\} $) are
adaptive to the instantaneous cache state $S$, CSI $\mathbf{H}$
and QSI $\mathbf{Q}$ to exploit the opportunistic CoMP gain and to
guarantee the QoS requirements of the users for given $\mathbf{q}$
and $\pi$. The proposed two timescale control structure and the corresponding
algorithm components are illustrated in Fig. \ref{fig:moduleconn}.
The ZF beamforming component and the random cache state generator
have been discussed in Section \ref{sec:System-Model}. The power
control and cache control are two key algorithm components that will
be respectively elaborated in Section \ref{sec:Low-Complexity-Power}
and Section \ref{sec:Asymptotically-Optimal-out_Solution}. The other
algorithm components and the overall solution will be discussed in
Section \ref{sec:Implementation-Considerations}.

\subsection{Power Control Policy}

Let $\boldsymbol{\chi}=\left(\mathbf{Q},S,\mathbf{H}\right)$ denote
the global system state. At the beginning of each time slot, the BS
determines the power control actions based on the global system state
$\boldsymbol{\chi}$ according to the following stationary control
policy. 
\begin{defn}
[Stationary Power Control Policy]A stationary power control policy
for the $k$-th user $\Omega_{k}$ is a mapping from the global system
state $\boldsymbol{\chi}$ to the power control actions of the $k$-th
user. Specifically, we have $p_{k}=\Omega_{k}\left(\boldsymbol{\chi}\right)$.
Furthermore, let $\Omega=\left\{ \Omega_{k},\forall k\right\} $ denote
the aggregation of the control policies for all the $2M$ users.
\end{defn}

Given $\mathbf{q}$, $\pi$ and a control policy $\Omega$, the induced
random process $\boldsymbol{\chi}\left(t\right)$ is a controlled
Markov chain with the following transition probability:
\begin{eqnarray*}
\textrm{Pr}^{\left(\mathbf{q},\pi,\Omega\right)}\left[\boldsymbol{\chi}\left(t+1\right)\left|\boldsymbol{\chi}\left(t\right)\right.\right]=\;\;\;\;\;\;\;\;\;\;\;\;\;\;\;\;\;\;\;\;\;\;\;\;\;\;\;\;\;\;\;\;\;\;\;\;\;\;\;\;\;\;\;\;\;\;\;\\
\textrm{Pr}^{\left(\mathbf{q},\pi,\Omega\right)}\left[S\left(t+1\right)\right]\textrm{Pr}\left[\mathbf{H}\left(t+1\right)\right]\textrm{Pr}^{\left(\mathbf{q},\pi,\Omega\right)}\left[\mathbf{Q}\left(t+1\right)\left|\boldsymbol{\chi}\left(t\right)\right.\right],
\end{eqnarray*}
where $\textrm{Pr}^{\left(\mathbf{q},\pi,\Omega\right)}$ means the
probability w.r.t. the measure induced by the control policy $\Omega$
under given $\mathbf{q},\pi$; the queue transition probability is
given by
\begin{eqnarray*}
 &  & \textrm{Pr}^{\left(\mathbf{q},\pi,\Omega\right)}\left[\mathbf{Q}\left(t+1\right)\left|\boldsymbol{\chi}\left(t\right)\right.\right]\\
 & = & \begin{cases}
1, & \textrm{if}\: Q_{k}\left(t+1\right),\forall k\:\textrm{is\:\ given\:\ by}\:(\ref{eq:Qdyn}),\\
0, & \textrm{otherwise}.
\end{cases}
\end{eqnarray*}
The admissible control policy in this paper is defined as follows. 
\begin{defn}
[Admissible Control Policy]For given $\mathbf{q},\pi$, a policy
$\Omega$ is admissible if the following requirements are satisfied: 
\begin{enumerate}
\item $\Omega$ is a unichain policy, i.e., the controlled Markov chain
$\boldsymbol{\chi}\left(t\right)$ under $\Omega$ has a single recurrent
class (and possibly some transient states) \cite{Bertsekas_MIT07_DynProg}.
\item The queueing system under $\Omega$ is stable in the sense that $\underset{T\rightarrow\infty}{\textrm{lim}}\textrm{E}^{\left(\mathbf{q},\pi,\Omega\right)}\left[\sum_{k=1}^{K}Q_{k}\left(T\right)\right]<\infty$,
where $\textrm{E}^{\left(\mathbf{q},\pi,\Omega\right)}\left[\cdot\right]$
means taking expectation w.r.t. the probability measure induced by
the control policy $\Omega$ under given $\mathbf{q},\pi$.
\end{enumerate}
\end{defn}

\subsection{Video Streaming Performance and Cost\label{sub:Performance-Costs}}

\begin{figure}
\begin{centering}
\textsf{\includegraphics[clip,width=85mm]{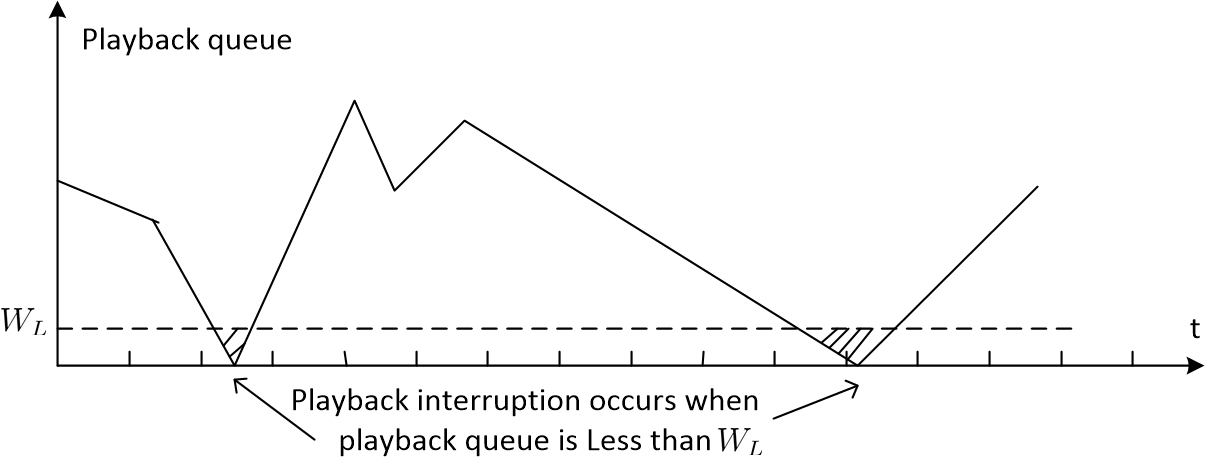}}
\par\end{centering}

\caption{\label{fig:plabackQueue}{\small{Illustration of the playback process
and playback interruption at the user.}}}
\end{figure}

For video streaming applications, there is a playback buffer at the
users as illustrated in Fig. \ref{fig:Playbuffer}. The output of
the playback buffer is the video decoding rate (playback rate) but
the input of the buffer is random (randomness induced by the channel
fading). During video streaming session, the user do not expect interruption
in the playback process. However, playback interruption event occurs
whenever the queue length of the playback buffer is less than $W_{L}$
as illustrated in Fig. \ref{fig:plabackQueue}. Hence, an important
first order performance metric for video streaming is the probability
of playback interruption. On the other hand, the playback buffer cost
in the mobile must be explicitly accounted for. As a result, the system
performance and cost are characterized by the \textit{interruption
probability}, the \textit{playback buffer overflow probability} of
the users (user centric performance metrics) and the average transmit
power at the BS and RS (network centric performance metric).

For given $\mathbf{q},\pi$ and an admissible control policy $\Omega$,
the interruption probability and the playback buffer overflow probability
of the $k$-th user are given by
\begin{eqnarray}
I_{k}^{\left(\mathbf{q},\pi,\Omega\right)} & = & \underset{T\rightarrow\infty}{\textrm{lim sup}}\frac{1}{T}\sum_{t=1}^{T}\textrm{E}^{\left(\mathbf{q},\pi,\Omega\right)}\left[1\left(Q_{k}\left(t\right)<W_{L}\right)\right]\label{eq:defI}\\
B_{k}^{\left(\mathbf{q},\pi,\Omega\right)} & = & \underset{T\rightarrow\infty}{\textrm{lim sup}}\frac{1}{T}\sum_{t=1}^{T}\textrm{E}^{\left(\mathbf{q},\pi,\Omega\right)}\left[1\left(Q_{k}\left(t\right)>W_{H}\right)\right]\label{eq:defB}
\end{eqnarray}
where $W_{H}>W_{L}$ is the target maximum playback buffer size at
the mobile. For technical reasons, we respectively use $e^{-\alpha\left(Q_{k}-W_{L}\right)^{+}}$
and $e^{-\alpha\left(W_{H}-Q_{k}\right)^{+}}$ as a smooth approximation
of the indicator function in (\ref{eq:defI}) and (\ref{eq:defB}),
where $\alpha>0$ is a smoothing parameter. As a result, the smooth
approximation of the \textit{interruption probability} and \textit{overflow
probability} is given by:
\begin{eqnarray}
\overline{I}_{k}^{\left(\mathbf{q},\pi,\Omega\right)} & = & \underset{T\rightarrow\infty}{\textrm{lim sup}}\frac{1}{T}\sum_{t=1}^{T}\textrm{E}^{\left(\mathbf{q},\pi,\Omega\right)}\left[e^{-\alpha\left(Q_{k}\left(t\right)-W_{L}\right)^{+}}\right]\label{eq:defI-1}\\
\overline{B_{k}}^{\left(\mathbf{q},\pi\Omega\right)} & = & \underset{T\rightarrow\infty}{\textrm{lim sup}}\frac{1}{T}\sum_{t=1}^{T}\textrm{E}^{\left(\mathbf{q},\pi,\Omega\right)}\left[e^{-\alpha\left(W_{H}-Q_{k}\left(t\right)\right)^{+}}\right]\label{eq:defB-1}
\end{eqnarray}
It can be shown that the smooth approximation is tight ($\overline{I}_{k}^{\left(\mathbf{q},\pi,\Omega\right)}\rightarrow I_{k}^{\left(\mathbf{q},\pi,\Omega\right)}$
and $\overline{B_{k}}^{\left(\mathbf{q},\pi\Omega\right)}\rightarrow B_{k}^{\left(\mathbf{q},\pi,\Omega\right)}$)
as $\alpha\rightarrow\infty$. The average total transmit power at
the BS and RS is given by: 
\[
\overline{P}^{\left(\mathbf{q},\pi,\Omega\right)}=\underset{T\rightarrow\infty}{\textrm{lim sup}}\frac{1}{T}\sum_{t=1}^{T}\textrm{E}^{\left(\mathbf{q},\pi,\Omega\right)}\left[\sum_{k=1}^{2M}p_{k}\left(t\right)\right].
\]
Note that the cost related to the RS is reflected in the average transmit
power cost $\overline{P}^{\left(\mathbf{q},\pi,\Omega\right)}$, which
is a summation of the average transmit power at the BS and the average
transmit power at the RS. As a result, the final system cost is given
by:
\begin{eqnarray*}
 &  & \overline{C}\left(\mathbf{q},\pi,\Omega\right)\\
 & = & \sum_{k=1}^{2M}\left(\beta_{k}\overline{I}_{k}^{\left(\mathbf{q},\pi,\Omega\right)}+\gamma_{k}\overline{B_{k}}^{\left(\mathbf{q},\pi,\Omega\right)}\right)+\overline{P}^{\left(\mathbf{q},\pi,\Omega\right)}\\
 & = & \underset{T\rightarrow\infty}{\textrm{lim sup}}\frac{1}{T}\sum_{t=1}^{T}\textrm{E}^{\left(\mathbf{q},\pi,\Omega\right)}\left[c\left(\mathbf{Q}\left(t\right),\Omega\left(\boldsymbol{\chi}\left(t\right)\right)\right)\right],
\end{eqnarray*}
where $\beta_{k}>0$ and $\gamma_{k}>0$ are respectively the interruption
price and buffer price of user $k$; $c\left(\mathbf{Q},\Omega\left(\boldsymbol{\chi}\right)\right)=\sum_{k=1}^{2M}\left(\hat{c}_{k}\left(Q_{k}\right)+p_{k}\right)$
with
\begin{equation}
\hat{c}_{k}\left(Q_{k}\right)=\beta_{k}e^{-\alpha\left(Q_{k}-W_{L}\right)^{+}}+\gamma_{k}e^{-\alpha\left(W_{H}-Q_{k}\right)^{+}}.\label{eq:cheadQ}
\end{equation}
We have the following additional assumptions on $\beta_{k},\gamma_{k}$
and $\alpha$.

\begin{assumption}[Assumptions on the Price Parameters $\left(\beta_{k},\gamma_{k}\right)$]\label{asm:pricepara}The
parameters $\beta_{k},\gamma_{k}\forall k$ and $\alpha$ satisfy
\begin{eqnarray}
 &  & e^{\left(W_{L}-W_{H}\right)\alpha}<\frac{\beta_{k}}{\gamma_{k}}<e^{\left(W_{H}-W_{L}\right)\alpha},\:\forall k,\label{eq:PullQ0}\\
 &  & \beta_{k}>\frac{\frac{1}{2}\left(\lambda_{0}e^{-\frac{1}{\lambda_{0}}}-E_{1}\left(\frac{1}{\lambda_{0}}\right)\right)+\gamma_{k}e^{-\frac{\left(W_{H}-W_{L}\right)\alpha}{2}}}{1-e^{-\frac{\left(W_{H}-W_{L}\right)\alpha}{2}}},\label{eq:beta0}
\end{eqnarray}
where $E_{1}\left(x\right)=\int_{x}^{\infty}\frac{e^{-t}}{t}dt$;
$\lambda_{0}$ is the unique solution of $\frac{B_{W}}{2\textrm{ln}2}E_{1}\left(1/\lambda\right)=\mu_{0}$
w.r.t. $\lambda>0$.\end{assumption}

If $\beta_{k}$ is too small, the power cost will dominate and the
users will not be served at all (i.e., they are always allocated with
zero power). The condition in (\ref{eq:beta0}) is used to avoid such
an uninteresting degenerate case. Define
\begin{equation}
Q_{k}^{\circ}\triangleq\underset{Q_{k}}{\textrm{min}}\:\hat{c}_{k}\left(Q_{k}\right).\label{eq:Q0def}
\end{equation}
The condition in (\ref{eq:PullQ0}) is used to ensure that the per
stage cost function $\hat{c}_{k}\left(Q_{k}\right)$ of user $k$
(excluding the power cost) attains the minimum at the queue length
$Q_{k}^{\circ}\in\left(W_{L},W_{H}\right)$ which corresponds to the
\textquotedbl{}desired\textquotedbl{} operating regime. Mathematically,
Assumption \ref{asm:pricepara} ensures that the solution of the approximated
Bellman equation in Theorem \ref{thm:Per-user-Bellman-Solution} is
valid. We also give an intuitive interpretation for the conditions
in Assumption \ref{asm:pricepara}. Suppose that the queueing system
$Q_{k}\left(t\right)$ is concentrated around $Q_{k}^{\Omega}$ under
the control policy $\Omega$. Then the corresponding average arrival
rate $\textrm{E}^{\left(\mathbf{q},\pi,\Omega\right)}\left[r_{k}\left(\mathbf{g}_{k},\Omega_{k}\left(\mathbf{Q},S,\mathbf{H}\right)\right)\right]\approx\mu_{k}\left(Q_{k}^{\Omega}\right)$.
Hence the per user average cost under $\Omega$ can be approximated
as $\hat{\overline{C}}_{k}\left(Q_{k}^{\Omega}\right)=\hat{c}_{k}\left(Q_{k}^{\Omega}\right)+\hat{\overline{P}}_{k}\left(Q_{k}^{\Omega}\right)$,
where $\hat{\overline{P}}_{k}\left(Q_{k}^{\Omega}\right)$ is the
average power required to support an average rate of $\mu_{k}\left(Q_{k}^{\Omega}\right)$.
It follows from $\mu_{k}\left(Q_{k}\right)=\mu_{0},\forall Q_{k}\in\left[W_{L},\infty\right)$
that $\hat{\overline{C}}_{k}\left(Q_{k}^{\circ}\right)<\hat{\overline{C}}_{k}\left(Q_{k}\right),\forall Q_{k}\neq Q_{k}^{\circ}\in\left[W_{L},\infty\right)$.
Hence the optimal control policy $\Omega^{*}$ that minimizes $\hat{\overline{C}}_{k}\left(Q_{k}^{\Omega}\right)$
either satisfies $Q_{k}^{\Omega^{*}}<W_{L}$ or $Q_{k}^{\Omega^{*}}=Q_{k}^{\circ}$.
To avoid the uninteresting degenerate case where $Q_{k}^{\Omega^{*}}<W_{L}$,
we require $\hat{\overline{C}}_{k}\left(Q_{k}\right)>\hat{\overline{C}}_{k}\left(Q_{k}^{\circ}\right),\forall Q_{k}\in\left[0,W_{L}\right]$.
Using the fact that $\hat{\overline{C}}_{k}\left(Q_{k}\right)\geq\beta_{k},\forall Q_{k}\in\left[0,W_{L}\right]$,
$\hat{c}_{k}\left(Q_{k}^{\circ}\right)\leq\hat{c}_{k}\left(\frac{W_{H}-W_{L}}{2}\right)$
and $\hat{\overline{P}}_{k}\left(Q_{k}^{\circ}\right)\leq\frac{\lambda_{0}e^{-\frac{1}{\lambda_{0}}}-E_{1}\left(\frac{1}{\lambda_{0}}\right)}{2}$
(this is because $\frac{\lambda_{0}e^{-\frac{1}{\lambda_{0}}}-E_{1}\left(\frac{1}{\lambda_{0}}\right)}{2}$
is the average power required to support an average rate of $\mu_{k}\left(Q_{k}^{\circ}\right)=\mu_{0}$
when $q_{\textrm{min}}=0$), it is easy to see that the condition
in (\ref{eq:beta0}) ensures that $\hat{\overline{C}}_{k}\left(Q_{k}\right)>\hat{\overline{C}}_{k}\left(Q_{k}^{\circ}\right),\forall Q_{k}\in\left[0,W_{L}\right]$.

\subsection{Two timescale Optimization Formulation}

The joint cache and power control problem is formulated as the following
two timescale optimization problem:
\[
\mathcal{P}:\:\underset{\mathbf{q}}{\textrm{min}}\:\textrm{E}\left[\underset{\Omega}{\textrm{min}}\:\overline{C}\left(\mathbf{q},\pi,\Omega\right)\right]+\eta\sum_{l=1}^{L}F_{l}q_{l},\:\textrm{s.t}.\: q_{l}\in\left[0,1\right],\forall l,
\]
where the expectation is taken w.r.t. the distribution of $\pi$;
the term $\sum_{l=1}^{L}F_{l}q_{l}$ represents the cache cost at
the RS; and $\eta>0$ is the cache price.
\begin{rem}
[Interpretation of Pricing Vector]The focus of the paper is to solve
\textbf{$\mathcal{P}$} for a given pricing vector \textbf{$\left[\beta_{1},...,\beta_{2M},\gamma_{1},...,\gamma_{2M},\eta\right]$}.\textbf{
}In fact, the vector \textbf{$\left[\beta_{1},...,\beta_{2M},\gamma_{1},...,\gamma_{2M},\eta\right]$}
can be interpreted as the \textit{pricing vectors} or the \textit{Lagrange
Multipliers} (LM). 
\begin{itemize}
\item \textbf{Pricing Vector Interpretation: }In practice, a Pareto optimal
solution%
\footnote{There are three different types of costs in video streaming, namely,
the interruption cost of each user, the playback buffer cost of each
user and the average total transmit power cost. A solution is called
Pareto optimal if it is impossible to decrease one cost without increasing
at least one of the other costs.%
} from the knee region of the Pareto front is usually more preferable
than other solutions \cite{Rachmawati_TEC09_MOEA}. Once we can solve
problem $\mathcal{P}$ (for a given pricing vector), there are existing
algorithms (such as multi-objective evolutionary algorithm (MOEA)
\cite{Rachmawati_TEC09_MOEA,Poor_TOCS13_MOPNet}) that can be used
to determine the appropriate pricing vector to achieve the knee solutions.
\item \textbf{LM Vector Interpretation:} In practice, we usually have a
constraint on the RS cache size. The cache price $\eta$ can be interpreted
as the Lagrange multiplier associated with the RS cache size constraint.
Similarly, the prices $\beta_{k}$ and $\gamma_{k}$ can be interpreted
as the Lagrange multipliers associated with the constraints on interruption
probability and playback buffer overflow probability requirements
for the $k$-th video streaming user. 
\end{itemize}
\end{rem}

Using primal decomposition \cite{Boyd_03note_primaldecomp}, problem
$\mathcal{P}$ can be decomposed into two subproblems.

\textbf{Inner MDP Subproblem }(\textbf{\small{Short-term Power Control}}):
Optimization of $\Omega$ for given $\mathbf{q},\pi$. 
\[
\mathcal{P}_{I}\left(\mathbf{q},\pi\right):\:\overline{C}^{*}\left(\mathbf{q},\pi\right)\triangleq\underset{\Omega}{\textrm{min}}\:\overline{C}\left(\mathbf{q},\pi,\Omega\right).
\]

\textbf{Outer Caching Subproblem }(\textbf{\small{Long-term Cache
Control}}):\textbf{ }Optimization of $\mathbf{q}$.
\[
\mathcal{P}_{O}:\:\underset{\mathbf{q}}{\textrm{min}}\:\textrm{E}\left[\overline{C}^{*}\left(\mathbf{q},\pi\right)\right]+\eta\sum_{l=1}^{L}F_{l}q_{l},\:\textrm{s.t}.\: q_{l}\in\left[0,1\right],\forall l.
\]

Subproblem $\mathcal{P}_{I}\left(\mathbf{q},\pi\right)$ belongs to
the infinite horizon average cost MDP, and it is well known that there
is no simple solution associated with such MDP. Brute force value
iterations or policy iterations \cite{Bertsekas_MIT07_DynProg} could
not lead to any viable solutions due to the curse of dimensionality.
Subproblem $\mathcal{P}_{O}$ is also a difficult stochastic optimization
problem. The term $\overline{C}^{*}\left(\mathbf{q},\pi\right)$ in
the objective function depends on the optimal solution of the inner
MDP problem and there is no closed form characterization. Moreover,
the distribution of $\pi$ is unknown and we cannot calculate the
expectation $\textrm{E}\left[\overline{C}^{*}\left(\mathbf{q},\pi\right)\right]$
explicitly. In Section \ref{sec:Low-Complexity-Power}, by exploiting
the special structure in our problem, we derive an approximate Bellman
equation to obtain a low complexity solution for the inner MDP problem
$\mathcal{P}_{I}\left(\mathbf{q},\pi\right)$. In Section \ref{sec:Asymptotically-Optimal-out_Solution},
we derive a closed form convex approximation of $\overline{C}^{*}\left(\mathbf{q},\pi\right)$,
which is the key to develop a robust stochastic subgradient algorithm
to find an asymptotically optimal solution for $\mathcal{P}_{O}$.

\section{Low Complexity Power Control Solution for $\mathcal{P}_{I}\left(\mathbf{q},\pi\right)$\label{sec:Low-Complexity-Power}}

To simplify notation, we use $\textrm{E}\left[\cdot\right]$ to denote
$\textrm{E}^{\left(\mathbf{q},\pi,\Omega\right)}\left[\cdot\right]$,
and $\textrm{Pr}\left[\cdot|\Omega\right]$ to denote $\textrm{Pr}^{\left(\mathbf{q},\pi,\Omega\right)}\left[\cdot\right]$
in this section.

\subsection{Optimality Conditions and Approximate Bellman Equation}

Utilizing the i.i.d. property of $\left\{ S,\mathbf{H}\right\} $
w.r.t. time slots, the following gives the optimality condition of
problem $\mathcal{P}_{I}\left(\mathbf{q},\pi\right)$. 
\begin{thm}
[Sufficient Optimality Conditions for $\mathcal{P}_{I}\left(\mathbf{q},\pi\right)$]\label{thm:Equ_bellman}For
given $\mathbf{q},\pi$, assume there exists a $\left(\theta^{*},\left\{ V^{*}\left(\mathbf{Q}\right)\right\} \right)$
that solves the following equivalent Bellman equation:{\small{
\begin{eqnarray}
\theta^{*}\tau+V^{*}\left(\mathbf{Q}\right) & = & \textrm{E}\bigg[\underset{\left\{ p_{k}\right\} }{\textrm{min}}\bigg[c\left(\mathbf{Q},\left\{ p_{k}\right\} \right)\tau+\nonumber \\
 &  & \sum_{\mathbf{Q}^{'}}\textrm{Pr}\left[\mathbf{Q}^{'}|\boldsymbol{\chi},\left\{ p_{k}\right\} \right]V^{*}\left(\mathbf{Q}^{'}\right)\bigg]\bigg|\mathbf{Q}\bigg]\nonumber \\
 &  & \forall\mathbf{Q}\in\boldsymbol{\mathcal{Q}},\label{eq:Equ_Bellman}
\end{eqnarray}
}}Furthermore, for all admissible control policy $\Omega$ and initial
queue state $\mathbf{Q}\left(0\right)$, $V^{*}$ satisfies the following
transversality condition:
\begin{equation}
\underset{T\rightarrow\infty}{\textrm{lim sup}}\frac{1}{T}\textrm{E}\left[\left.V^{*}\left(\mathbf{Q}\left(T\right)\right)\right|\mathbf{Q}\left(0\right)\right]=0.\label{eq:transcond}
\end{equation}
Then, $\theta^{*}=\overline{C}^{*}\left(\mathbf{q},\pi\right)$ is
the optimal average cost and $V^{*}\left(\mathbf{Q}\right)$ is called
the relative value function. If $\left\{ p_{k}^{*}\right\} $ attains
the minimum of the R.H.S. of (\ref{eq:Equ_Bellman}) for given $\boldsymbol{\chi}$,
then $\Omega^{*}\left(\boldsymbol{\chi}\right)=\left\{ p_{k}^{*}\right\} $
is the optimal control policy for Problem $\mathcal{P}_{I}\left(\mathbf{q},\pi\right)$.
\end{thm}

Please refer to Appendix \ref{sub:Proof-of-TheoremEqubellman} for
the proof.

Note that the solution given by Theorem \ref{thm:Equ_bellman} is
unique due to the unichain assumption of the control policy \cite{Bertsekas_MIT07_DynProg}. 

Using Taylor expansion, we establish the following corollary on the
approximation of the Bellman equation%
\footnote{In smart grid applications, there are also stochastic optimization
problems to adapt the power generation according to dynamic loads.
The proposed approximate Bellman equation approach may potentially
be applied to solve these problems as well.%
} in (\ref{eq:Equ_Bellman}).
\begin{cor}
[Approximate Bellman Equation]\label{cor:Approximate-Bellman-EquationFor}For
any given $\mathbf{q},\pi$, if 
\begin{enumerate}
\item there is a unique $\left(\theta^{*},\left\{ V^{*}\left(\mathbf{Q}\right)\right\} \right)$
that satisfies the equivalent Bellman equation (\ref{eq:Equ_Bellman})
and transversality condition (\ref{eq:transcond}) in Theorem \ref{thm:Equ_bellman}.
\item there exist $\theta$ and $V\left(\mathbf{Q}\right)\in\mathcal{C}^{1}\left(\mathbb{R}_{+}^{2M}\right)$
that solve the following approximate Bellman equation%
\footnote{A function $f(\mathbf{x}),\mathbf{x}\in\mathbb{R}_{+}^{2M}$ belongs
to $\mathcal{C}^{1}\left(\mathbb{R}_{+}^{2M}\right)$ if the first
order partial derivative of $f(\mathbf{x})$ w.r.t. each element of
$\mathbf{x}$ is continuous.%
}:
\begin{eqnarray}
\theta=\textrm{E}\bigg[\underset{\left\{ p_{k}\right\} }{\textrm{min}}\bigg[c\left(\mathbf{Q},\left\{ p_{k}\right\} \right)+\sum_{k=1}^{2M}\frac{\partial V\left(\mathbf{Q}\right)}{\partial Q_{k}}\;\;\;\;\;\;\;\;\;\;\;\;\;\;\nonumber \\
\left.\times\left(r_{k}\left(\mathbf{g}_{k},p_{k}\right)-\mu_{k}\left(Q_{k}\right)\right)\bigg]\right|\mathbf{Q}\bigg],\label{eq:approxACOE}
\end{eqnarray}
for all $\mathbf{Q}\in\boldsymbol{\mathcal{Q}}$. Furthermore, for
all admissible control policy $\Omega$ and initial queue state $\mathbf{Q}(0)$,
the transversality condition in (\ref{eq:transcond}) is satisfied
for $V$.
\end{enumerate}
Then we have
\begin{eqnarray*}
\theta^{*} & = & \theta+o\left(1\right)\\
V^{*}\left(\mathbf{Q}\right) & = & V\left(\mathbf{Q}\right)+o\left(1\right),\:\forall\mathbf{Q}\in\boldsymbol{\mathcal{Q}},
\end{eqnarray*}
where the error term $o(1)$ asymptotically goes to zero for sufficiently
small slot duration $\tau$.
\end{cor}

Please refer to Appendix \ref{sub:Proof-of-CorollaryApproxACOE} for
the proof.

By Corollary \ref{cor:Approximate-Bellman-EquationFor}, the solution
of the Bellman equation (\ref{eq:Equ_Bellman}) can be approximated
by the solution of the approximate Bellman equation in (\ref{eq:approxACOE}),
and the approximation is asymptotically accurate as $\tau\rightarrow0$.
However, solving the approximate Bellman equation still involves solving
a large system of nonlinear fixed point equations. In the following
subsection, we exploit the specific structure of our problem to derive
a simple solution for the approximate Bellman equation.

\subsection{Solution for the Approximate Bellman Equation\label{sub:Solution-for-theBellman}}

In the proposed opportunistic CoMP scheme, the interference among
users is completely eliminated by ZF beamforming. As a result, the
dynamics of the playback buffer at the $2M$ users are decoupled and
we can obtain a closed form solution for the approximate Bellman equation
by solving a set of independent single-variable non-linear equations
as elaborated below.

We first derive the distribution of the effective channel $g_{k}$
for given $\mathbf{q}$ and $\pi$.
\begin{lem}
[PDF of the Effective Channel]\label{lem:PDF-of-EFC}For given $\mathbf{q}$
and $\pi$, the PDF of $g_{k}$ is given by
\begin{eqnarray*}
h_{g_{k}}\left(x\right) & = & \left(\frac{1+q_{\textrm{min}}}{2}\right)e^{-x}+\left(\frac{1-q_{\textrm{min}}}{2}\right)\delta\left(x\right),
\end{eqnarray*}
where $q_{\textrm{min}}=\underset{1\leq k\leq K}{\textrm{min}}\left\{ q_{\pi_{k}}\right\} $;
and $\delta(\cdot)$ denote the Dirac delta function.
\end{lem}

Please refer to Appendix \ref{Proof-of-Lemma-PDFCH} for the proof.

Recall the definition of $Q_{k}^{\circ}$ in (\ref{eq:Q0def}). Note
that under Assumption \ref{asm:pricepara}, we have 
\begin{equation}
Q_{k}^{\circ}=\frac{\textrm{log}\beta_{k}-\textrm{log}\gamma_{k}}{2\alpha}+\frac{W_{L}+W_{H}}{2}\in\left(W_{L},W_{H}\right).\label{eq:Q0val}
\end{equation}
Let $\tilde{\lambda}_{k}\left(x\right)$ denote the unique solution
of the equation 
\begin{equation}
\frac{B_{W}\left(1+q_{\textrm{min}}\right)}{2\textrm{ln}2}E_{1}\left(-\frac{\textrm{ln}2}{\lambda B_{W}}\right)-\mu_{k}\left(x\right)=0,\label{eq:Lamtutadef}
\end{equation}
w.r.t. $\lambda\in\mathbb{R}_{-}$ for any $x\in\mathcal{Q}$. Then
the following theorem holds.
\begin{thm}
[Closed-Form Solution of (\ref{eq:approxACOE})]\label{thm:Per-user-Bellman-Solution}For
any $1\leq k\leq2M$, let {\small{
\begin{eqnarray}
\tilde{\theta}_{k} & = & \hat{c}_{k}\left(Q_{k}^{\circ}\right)-\frac{\left(1+q_{\textrm{min}}\right)}{2}\label{eq:Thetatuta}\\
 &  & \times\left(\frac{\tilde{\lambda}_{k}\left(Q_{k}^{\circ}\right)B_{W}}{\textrm{ln}2}e^{\frac{\textrm{ln}2}{\tilde{\lambda}_{k}\left(Q_{k}^{\circ}\right)B_{W}}}+E_{1}\left(-\frac{\textrm{ln}2}{\tilde{\lambda}_{k}\left(Q_{k}^{\circ}\right)B_{W}}\right)\right),\nonumber 
\end{eqnarray}
}}and consider the following set of independent nonlinear equations
w.r.t. $f_{k}$ for all $x\in\mathcal{Q}$:{\small{
\begin{eqnarray}
\hat{c}_{k}\left(x\right)+\left(\frac{B_{W}\left(1+q_{\textrm{min}}\right)}{2\textrm{ln}2}E_{1}\left(-\frac{\textrm{ln}2}{f_{k}B_{W}}\right)-\mu_{k}\left(x\right)\right)f_{k}\label{eq:ODEVQ}\\
-\frac{\left(1+q_{\textrm{min}}\right)}{2}\left(\frac{f_{k}B_{W}}{\textrm{ln}2}e^{\frac{\textrm{ln}2}{f_{k}B_{W}}}+E_{1}\left(-\frac{\textrm{ln}2}{f_{k}B_{W}}\right)\right) & = & \tilde{\theta}_{k}.\nonumber 
\end{eqnarray}
}}Then the following are true:
\begin{enumerate}
\item For $x\in\left[0,Q_{k}^{\circ}\right]$, (\ref{eq:ODEVQ}) has a unique
solution denoted by $\tilde{f}_{k}\left(x\right),\: x\in\left[0,Q_{k}^{\circ}\right]$
over the region $f_{k}\in\left(-\infty,\tilde{\lambda}_{k}\left(x\right)\right]$.
\item For $x\in\left(Q_{k}^{\circ},\infty\right)$, (\ref{eq:ODEVQ}) has
a unique solution denoted by $\tilde{f}_{k}\left(x\right),\: x\in\left(Q_{k}^{\circ},\infty\right)$
over the region $f_{k}\in\left[\tilde{\lambda}_{k}\left(x\right),\infty\right]$. 
\item $\tilde{\theta}\triangleq\sum_{k=1}^{2M}\tilde{\theta}_{k}$ and $\tilde{V}\left(\mathbf{Q}\right)\triangleq\sum_{k=1}^{2M}\int_{Q_{k}^{\circ}}^{Q_{k}}\tilde{f}_{k}\left(x\right)dx,\:\forall\mathbf{Q}\in\boldsymbol{\mathcal{Q}}$
satisfy the approximate Bellman equation (\ref{eq:approxACOE}). Moreover,
$\tilde{V}\left(\mathbf{Q}\right)$ satisfies the transversality condition
in (\ref{eq:transcond}) and thus $\theta^{*}=\tilde{\theta}+o\left(1\right)$
and $V^{*}\left(\mathbf{Q}\right)=\tilde{V}\left(\mathbf{Q}\right)+o\left(1\right),\:\forall\mathbf{Q}\in\boldsymbol{\mathscr{Q}}$,
where $\theta^{*}$ is the optimal average cost and $V^{*}\left(\mathbf{Q}\right)$
is the optimal value function as defined in Corollary \ref{cor:Approximate-Bellman-EquationFor}.
\end{enumerate}
\end{thm}

Please refer to Appendix \ref{sub:Proof-of-TheoremCFbellman} for
the proof.

Based on the closed form solution for the approximate Bellman equation
in Theorem \ref{thm:Per-user-Bellman-Solution}, the power control
solution for given QSI $\mathbf{Q}$, cache state $S$ and CSI $\mathbf{H}$
is
\begin{equation}
p_{k}=\left(-\frac{\tilde{f}_{k}\left(Q_{k}\right)B_{W}}{\textrm{ln}2}-\frac{1}{g_{k}}\right)^{+},\forall k.\label{eq:optpow}
\end{equation}
Note that the power control solution in (\ref{eq:optpow}) only requires
the partial derivatives of the approximate value function $\frac{\partial\tilde{V}\left(\mathbf{Q}\right)}{\partial Q_{k}}\triangleq\tilde{f}_{k}\left(Q_{k}\right),\forall k$,
which can be easily found by solving (\ref{eq:ODEVQ}) in Theorem
\ref{thm:Per-user-Bellman-Solution}. The solution in (\ref{eq:optpow})
has a multi-level water-filling structure, where the water level depends
on the QSI via $\tilde{f}_{k}\left(Q_{k}\right)$, which captures
the urgency of the $k$-th data flow. The overall power control algorithm
is summarized in Table \ref{tab:Algpow}. By Theorem \ref{thm:Per-user-Bellman-Solution},
the proposed power control algorithm is asymptotically optimal as
$\tau\rightarrow0$. Furthermore, the cache underflow probability
under the proposed power control is negligible for large $L_{S}$
as proved in the following proposition.
\begin{prop}
[Cache Underflow for Large $L_{S}$]\label{prop:cacheunderflow}For
given $\mathbf{q},\pi$ and under the power control policy in Table
\ref{tab:Algpow}, the cache underflow probability tends to zero as
the segment size $L_{S}\rightarrow\infty$.
\end{prop}

Please refer to Appendix \ref{sub:Proof-of-PropositionUF} for the
proof.
\begin{table}
\caption{\label{tab:Algpow}Algorithm I (Power control for Problem $\mathcal{P}_{I}\left(\mathbf{q},\pi\right))$}

\centering{}%
\begin{tabular}{l}
\hline 
\textbf{\small{Actions at the beginning of each time slot:}}{\small{ }}\tabularnewline
{\small{$\:$$\:$1: Calculate $\tilde{f}_{k}\left(Q_{k}\right),\forall k$
by solving (\ref{eq:ODEVQ}) with current $\mathbf{Q}$.}}\tabularnewline
{\small{$\:$$\:$2: Obtain the transmit powers using (\ref{eq:optpow})
with current $g_{k},\forall k$.}}\tabularnewline
\hline 
\end{tabular}
\end{table}

\section{Asymptotically Optimal Solution for $\mathcal{P}_{O}$\label{sec:Asymptotically-Optimal-out_Solution}}

One challenge of solving $\mathcal{P}_{O}$ is that the objective
function depends on the optimal average cost $\overline{C}^{*}\left(\mathbf{q},\pi\right)$
of the inner MDP problem, which is difficult to obtain. To tackle
this difficulty, we derive an asymptotically accurate closed form
approximation for $\overline{C}^{*}\left(\mathbf{q},\pi\right)$.
\begin{thm}
[Asymptotic Approximation of $\overline{C}^{*}\left(\mathbf{q},\pi\right)$]\label{thm:Asymptotic-Approximation-of-Cbar}As
$\frac{\mu_{0}}{B_{W}}\rightarrow\infty$ and $\tau\rightarrow0$,
we have 
\[
\frac{\left|\overline{C}^{*}\left(\mathbf{q},\pi\right)-\hat{C}\left(\mathbf{q},\pi\right)\right|}{\overline{C}^{*}\left(\mathbf{q},\pi\right)}=\mathcal{O}\left(\frac{\mu_{0}}{B_{W}}e^{-\frac{\mu_{0}}{B_{W}}}\right)\rightarrow0,
\]
where
\[
\hat{C}\left(\mathbf{q},\pi\right)=\sum_{k=1}^{2M}\left(\xi e^{-a_{1}+\frac{\mu_{0}\textrm{ln}2}{B_{W}\xi}}-(a_{2}+e)\xi+e+\hat{c}_{k}\left(Q_{k}^{\circ}\right)\right),
\]
where $\xi=\frac{1+\underset{1\leq k\leq2M}{\textrm{min}}\left\{ q_{\pi_{k}}\right\} }{2}$;
$a_{1}=\textrm{log}\left(e^{-\frac{\mu_{0}\textrm{ln}2}{B_{W}}}\right)e^{-e^{-\frac{\mu_{0}\textrm{ln}2}{B_{W}}}}+E_{1}\left(e^{-\frac{\mu_{0}\textrm{ln}2}{B_{W}}}\right)$
and $a_{2}=E_{1}\left(e^{-\frac{\mu_{0}\textrm{ln}2}{B_{W}}}\right)$;
$\hat{c}_{k}\left(\cdot\right)$ and $Q_{k}^{\circ}$ are given in
(\ref{eq:cheadQ}) and (\ref{eq:Q0val}) respectively.
\end{thm}

Please refer to Appendix \ref{Proof-of-TheoremCbar} for the proof.

Then the following corollary follows immediately from Theorem \ref{thm:Asymptotic-Approximation-of-Cbar}.
\begin{cor}
[Asymptotic equivalence of $\mathcal{P}_{O}$]\label{cor:Asymptotic-equivalence-ofPO}Let
$\tilde{\mathbf{q}}$ be an optimal solution of the problem
\[
\mathcal{P}_{E}:\:\underset{\mathbf{q}}{\textrm{min}}\: U\left(\mathbf{q}\right),\:\textrm{s.t}.\: q_{l}\in\left[0,1\right],\forall l,
\]
where $U\left(\mathbf{q}\right)=\textrm{E}\left[\hat{C}\left(\mathbf{q},\pi\right)\right]+\eta\sum_{l=1}^{L}F_{l}q_{l}$.
Let $C^{*}$ be the optimal value of $\mathcal{P}_{O}$. Then we have
\begin{eqnarray}
\left|\textrm{E}\left[\overline{C}^{*}\left(\tilde{\mathbf{q}},\pi\right)\right]+\eta\sum_{l=1}^{L}F_{l}\tilde{q}_{l}-C^{*}\right|/C^{*} & \rightarrow & 0,\label{eq:TS1}
\end{eqnarray}
as $\frac{\mu_{0}}{B_{W}}\rightarrow\infty$ and $\tau\rightarrow0$.
\end{cor}

By Corollary \ref{cor:Asymptotic-equivalence-ofPO}, the solution
of $\mathcal{P}_{O}$ can be approximated by the solution of $\mathcal{P}_{E}$,
and the approximation is asymptotically accurate for high SNR%
\footnote{Note that a large $\frac{\mu_{0}}{B_{W}}$ implies high SNR.%
} and small slot size $\tau$. It can be shown that $\xi$ is a concave
function w.r.t. $\mathbf{q}$ and $\hat{C}\left(\mathbf{q},\pi\right)$
is decreasing with $\xi$ for $q_{l}\in\left[0,1\right],\forall l$.
Using the vector composition rule for convex function \cite{Boyd_04Book_Convex_optimization},
$\hat{C}\left(\mathbf{q},\pi\right)$ is convex w.r.t. $\mathbf{q}$
and thus $\mathcal{P}_{E}$ is also a convex problem. Hence, we propose
a stochastic subgradient algorithm which is able to converge to the
optimal solution of $\mathcal{P}_{E}$ without knowing the distribution
of $\pi$. The algorithm is summarized in Table \ref{tab:Algq} and
the global convergence is established in the following Theorem.
\begin{thm}
[Convergence of Algorithm E]If the step sizes $\sigma^{(i)}>0$
in Algorithm E satisfies: 1) $\sum_{i=1}^{\infty}\left(\sigma^{(i)}\right)^{2}<\infty$;
2) $\sum_{i=1}^{\infty}\sigma^{(i)}=\infty$, then Algorithm E converges
to an optimal solution $\tilde{\mathbf{q}}$ of Problem $\mathcal{P}_{E}$
with probability 1.
\end{thm}

The convergence proof follows directly from \cite[Theorem 3.3]{Ram_SIAM2009_SubgradConv}. 

\begin{table}
\caption{\label{tab:Algq}Algorithm E (for Solving Problem $\mathcal{P}_{E}$)}

\centering{}%
\begin{tabular}{l}
\hline 
\textbf{\small{Initialization: }}{\small{Choose initial $\mathbf{q}$:
$q_{l}\in\left[0,1\right],\forall l$. Let $i=1$.}}\tabularnewline
\textbf{\small{Step 1:}}{\small{ Calculate a noisy unbiased subgradient
of $U\left(\mathbf{q}\right)$,}}\tabularnewline
{\small{$\:$$\:$$\:$$\:$$\:$$\:$$\:$$\:$$\:$$\:$$\:$$\:$$\hat{\nabla}U=\left[\widehat{\frac{\partial U}{\partial q_{1}}},...,\widehat{\frac{\partial U}{\partial q_{L}}}\right]^{T}$
based on current $\mathbf{q},\pi$:}}\tabularnewline
{\small{$\widehat{\frac{\partial U}{\partial q_{l}}}=$}}\tabularnewline
$M\left[\left(1-\frac{\mu_{0}\textrm{ln}2}{B_{W}\xi}\right)e^{-a_{1}+\frac{\mu_{0}\textrm{ln}2}{B_{W}\xi}}-a_{2}-e\right]1\left(l=\pi_{k^{*}}\right)+\eta F_{l},\forall l,$\tabularnewline
{\small{where $k^{*}$ is any index satisfying $q_{\pi_{k^{*}}}=\underset{1\leq k\leq2M}{\textrm{min}}\left\{ q_{\pi_{k}}\right\} $. }}\tabularnewline
\textbf{\small{Step 2:}}{\small{ Choose proper step size $\sigma^{(i)}>0$
and update $\mathbf{q}$ as}}\tabularnewline
{\small{$\:$$\:$$\:$$\:$$\:$$q_{l}=\textrm{min}\left(\left(q_{l}-\sigma^{(i)}\widehat{\frac{\partial U}{\partial q_{l}}}\right)^{+},1\right),\forall l$}}\tabularnewline
\textbf{\small{Step 3:}}{\small{ Let $i=i+1$. When $\pi$ is changed,
return to Step 1.}}\tabularnewline
\hline 
\end{tabular}
\end{table}

\section{Implementation Considerations\label{sec:Implementation-Considerations}}

\subsection{Summary of the Overall Solution}

Fig. \ref{fig:moduleconn} summarizes the overall solution and the
inter-relationship of the algorithm components. The solutions are
divided into \textit{long timescale process} and \textit{short timescale
process}. The long timescale processing consists of Algorithm E (cache
control) and RS cache management. The short timescale processing consists
of random cache state generator, ZF beamforming and Algorithm I (power
control). The power control and cache control processes are implemented
at the BS, while the cache management process is implemented at the
RS. Whenever the URP $\pi$ changes, the updated cache control vector
$\mathbf{q}$ is computed from the BS and pass to the RS. Then the
RS cache management updates the RS cache according to $\mathbf{q}$.
Specifically, if the cached parity bits for each segment of the $l$-th
video file is less than $\frac{2q_{l}L_{S}}{1+q_{l}}$ bits, it will
request new parity bits from the low speed wireless backhaul. Otherwise,
it will drop some cached parity bits. At each time slot, the cache
state $S$ is generated from the random cache state generator using
$\mathbf{q}$ and $\pi$, the QSI $\mathbf{Q}$ is calculated at the
BS according to the queue dynamics in (\ref{eq:Qdyn}), and the CSI
$\mathbf{H}$ is obtained via feedback from the users. Furthermore,
at each time slot, the power control and ZF beamforming vectors are
determined at the BS based on the CSI $\mathbf{H}$, QSI $\mathbf{Q}$
and cache state $S$. If $S=1$, they are sent to the RS for opportunistic
CoMP transmission.

\subsection{Computational Complexity}

The main computation complexity of the proposed solution is dominated
by three algorithm components, namely, the ZF beamforming, the power
control Algorithm I and the long term cache control Algorithm E. We
analyze the complexity of each algorithm component as follows.
\begin{itemize}
\item \textbf{Complexity of ZF beamforming:} At each time slot, the calculation
of the ZF beamforming vectors $\overline{\mathbf{v}}_{k},\forall k\in\Theta$
(if $S=0$) requires a $M\times M$ matrix inversion and the calculation
of the ZF beamforming vectors $\mathbf{v}_{k},\forall k$ (if $S=1$)
requires a $2M\times2M$ matrix inversion. The complexity of ZF beamforming
is polynomial w.r.t. the number of antennas $M$ at the BS/RS. 
\item \textbf{Complexity of the power control Algorithm I:} At each time
slot, we need to calculate $\tilde{f}_{k}\left(Q_{k}\right),\forall k$
by solving $2M$ \textit{independent single-variable} nonlinear equations
as in (\ref{eq:ODEVQ}). Then the power control is calculated using
the multi-level water-filling solution in (\ref{eq:optpow}), which
only involves several floating point operations (FLOPs). Hence, the
complexity of Algorithm I only increases linearly with the number
of users $2M$.
\item \textbf{Complexity of the long term cache control Algorithm E:} For
each realization of $\pi$, we need to calculate the noisy unbiased
subgradient $\hat{\nabla}U$ of $U\left(\mathbf{q}\right)$. The subgradient
$\hat{\nabla}U$ has $L$ elements and the calculation of each element
$\widehat{\frac{\partial U}{\partial q_{l}}}$ only requires several
FLOPs. Then the cache control vector $\mathbf{q}$ is updated using
a simple subgradient method. Hence, the complexity of Algorithm E
only increases linearly with the number of video files $L$. Moreover,
the above subgradient update is performed at a much slower time scale
compared to the time slot rate.
\end{itemize}

In summary, using the closed-form approximation of $V^{*}\left(\mathbf{Q}\right)$
in Theorem 2, the proposed solution only has polynomial complexity
w.r.t. the number of antennas $M$ at the BS/RS, and linearly complexity
w.r.t. the number of users $2M$ and the number of video files $L$.
The complexity is very low compared with brute-force value iteration
\cite{Bertsekas_MIT07_DynProg} which has exponential complexity w.r.t.
$2M$. Table \ref{tab:Cputime} illustrates a comparison on the matlab
computational time among the brute force MDP, the baselines defined
in Section VII and the proposed solution. It can be seen that the
per time slot computation time of the proposed solution is similar
to the simple baselines without RS and is much lower than the decode
and forward (DF) relay scheme in \cite{Simoens_TSP09_MIMOrelay} or
the brute-force value iteration \cite{Bertsekas_MIT07_DynProg}. This
shows that the proposed solution is suitable for video streaming applications
that requires real-time processing.
\begin{table}
\centering{}{\small{\caption{\label{tab:Cputime}{\small{Comparison of the MATLAB computational
time of the proposed scheme and the baselines in one time slot. The
simulation setup is given in Section \ref{sec:Simulation-Results}.}}}
}}{\footnotesize{}}%
\begin{tabular}{|l|l|l|l|}
\hline 
 & {\small{$M=2$}} & {\small{$M=4$}} & {\small{$M=8$}}\tabularnewline
\hline 
{\small{Proposed}} & {\small{0.45ms}} & {\small{0.83ms}} & {\small{1.67ms}}\tabularnewline
{\small{Baseline 1,2 (without RS)}} & {\small{0.13ms}} & {\small{0.21ms}} & {\small{0.44ms}}\tabularnewline
{\small{Baseline 3 (Q-weighted DF)}} & {\small{2.12ms}} & {\small{4.56ms}} & {\small{10.42ms}}\tabularnewline
{\small{Brute-Force Value Iteration}} & 879s & $>10^{4}$s & $>10^{4}$s\tabularnewline
\hline 
\end{tabular}
\end{table}

\subsection{Signaling Overheads}

The short term signaling overhead is small compared to the conventional
CoMP schemes \cite{somekh2009cooperative,Irmer_Comm11_CoMPsurvey}.
At each time slot, the BS first broadcasts the cache state $S$ and
the user selection set $\Theta$ to the users. Then each selected
user $k$ feedbacks its channel vector $\overline{\mathbf{h}}_{k}$
(if $S=0$) or $\mathbf{h}_{k}$ (if $S=1$) to the BS. Finally, if
$S=1$, the BS sends the power allocation $\left\{ p_{k},\forall k\right\} $
and the beamforming vectors $\left\{ \mathbf{v}_{k},\forall k\right\} $
to the RS. The above signaling overhead is similar to the conventional
closed-loop multi-user MIMO systems \cite{Gesbert_SPM07_MUMIMO} and
can be supported by the modern wireless systems such as LTE \cite{LTE}.
The long term cache control signaling between the BS and RS is very
small since $\mathbf{q}$ is only sent to the RS once for each realization
of $\pi$. For the cache update process, whenever $q_{l}$ is increased,
the RS needs to request new parity bits from the wireless backhaul.
Since $\mathbf{q}$ is adaptive to the distribution of $\pi$ and
it changes very slowly, the wireless backhaul is enough to support
the cache update process. Table. \ref{tab:cacheupdateload} illustrates
the cache update loading (in terms of the average wireless backhaul
throughput used for cache update) versus different RS cache occupancy
(i.e., the total size of the cached data at RS). The corresponding
SNR gain over the baselines in Section \ref{sec:Simulation-Results}
(when achieving the same interruption probability $10^{-3}$) is also
illustrated.
\begin{table}
\centering{}\caption{\label{tab:cacheupdateload}{\small{Cache update loading under different
RS cache occupancy and SNR gains. We assume that the popularity of
the video files (i.e., the distribution of $\pi$) changes every week.
The system setup is given in Section \ref{sec:Simulation-Results}.
As a comparison, if conventional CoMP }}\cite{Irmer_Comm11_CoMPsurvey}{\small{
is employed, the backhaul loading between the BS and RS for exchanging
the payload data is $8$Mbps. }}}
{\footnotesize{}}%
\begin{tabular}{|l|l|l|l|}
\hline 
{\small{RS cache}} & {\small{Cache update}} & {\small{SNR gain over }} & {\small{SNR gain over }}\tabularnewline
{\small{occupancy}} & {\small{loading}} & {\small{Baseline 2}} & {\small{baseline 3}}\tabularnewline
\hline 
{\small{1.8G Bytes}} & {\small{25Kbps}} & {\small{4.9dB}} & {\small{3.8dB}}\tabularnewline
{\small{1.3G Bytes}} & {\small{18Kbps}} & {\small{3.7dB}} & {\small{2.6dB}}\tabularnewline
{\small{0.9G Bytes}} & {\small{12Kbps}} & {\small{2.6dB}} & {\small{1.5dB}}\tabularnewline
\hline 
\end{tabular}
\end{table}

\section{Simulation Results\label{sec:Simulation-Results}}

Consider a video streaming system with $L=6$ video files and $2M=4$
users. The size of each video file is $600$M Bytes and the streaming
rate is $2$M bits/s. Both the BS and RS are equipped with $M=2$
antennas. Assume that each user independently accesses the $l$-th
video file with probability $\rho_{l}$, and we set $\boldsymbol{\rho}=\left[\rho_{1},...,\rho_{6}\right]=\left[0.6,0.3,0.08,0.01,0.005,0.005\right]$,
which represents the popularity of different video files. Note that
$\boldsymbol{\rho}$ is only used to generate the realizations of
URP $\pi$ and the control algorithms do not have the knowledge of
$\boldsymbol{\rho}$. The other system parameters are set as%
\footnote{Note that although $\alpha$ is small, the approximations of the interruption
probability and playback buffer overflow probability in (\ref{eq:defI-1})
and (\ref{eq:defB-1}) are still good if $W_{H}\alpha$ is large as
will be shown in the simulations. %
}
\begin{eqnarray*}
B_{W}=1\textrm{MHz}, & \tau=5\textrm{ms}, & \alpha=7.5\times10^{-5},\\
W_{H}=2.5\times10^{5}\textrm{bits}, & W_{L}=2\times10^{4}\textrm{bits}.
\end{eqnarray*}

\subsection{Convergence of the Cache Control Algorithm E}

\begin{figure}
\begin{centering}
\includegraphics[width=85mm]{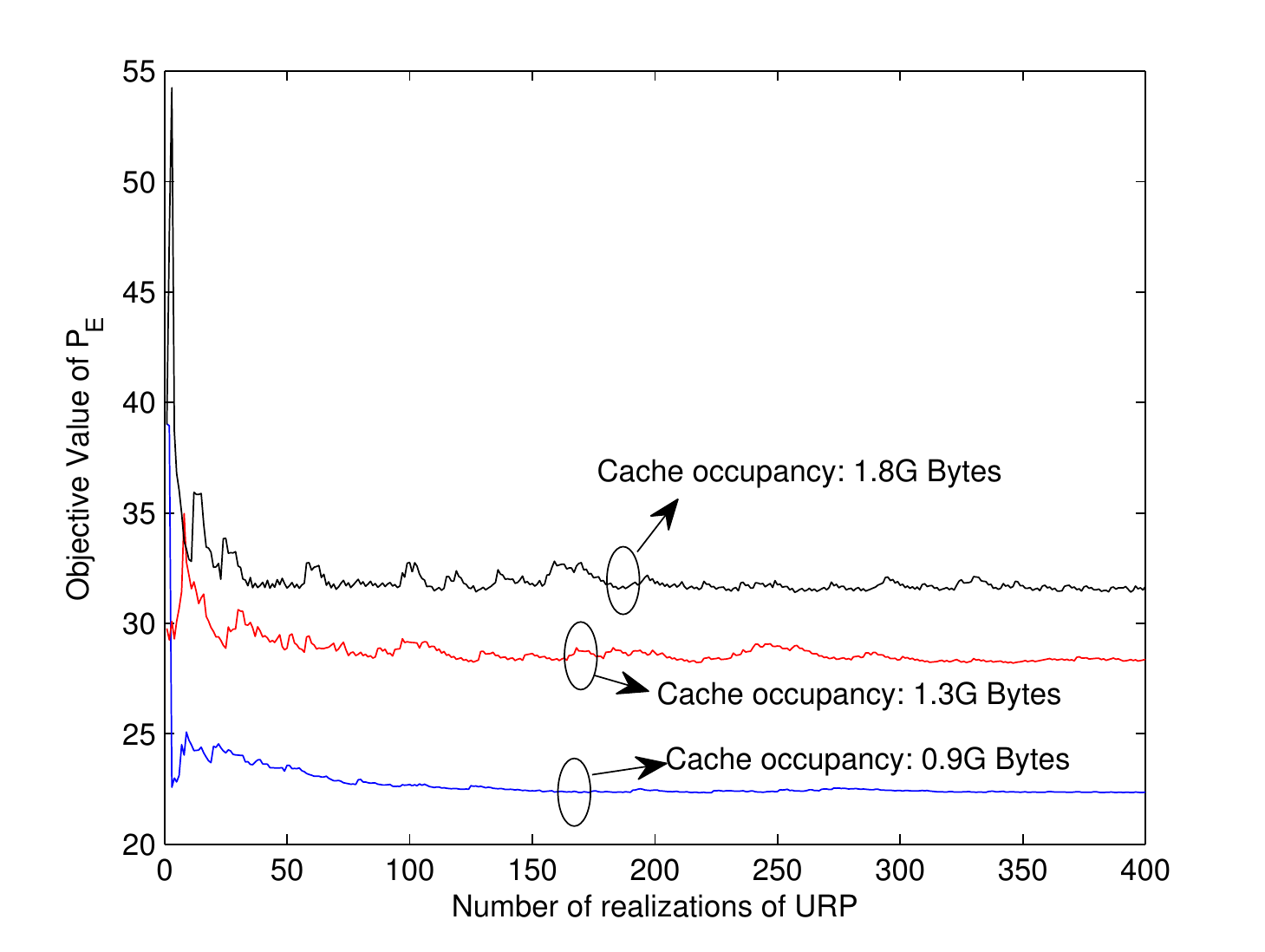}
\par\end{centering}

\caption{\label{fig:AlgE_conv}Convergence of Algorithm E}
\end{figure}

In Fig. \ref{fig:AlgE_conv}, we plot the objective value $U\left(\mathbf{q}\right)$
of $\mathcal{P}_{E}$ versus the number of realizations of $\pi$
(i.e., the number of iterations of Algorithm E) for different cache
prices $\eta$. We fix $\beta_{k}=\gamma_{k}=15,\:\forall k$. The
corresponding RS cache occupancy is also given besides each convergence
curve. It can be seen that Algorithm $E$ quickly converges.

\subsection{Performance Gain of the Proposed Solution w.r.t. Baselines}

The following baselines are considered.

\textbf{Baseline 1 (CSI-only without RS):} There is no RS. The physical
layer reduces to the Mode 0 ZF scheme in Section \ref{sub:Cache-enabled-Opportunistic-CoMP}.
The power control is only adaptive to the CSI. Specifically, at each
time slot, the power control is obtained by solving the following
optimization problem:
\[
\underset{\left\{ p_{k}\right\} }{\textrm{max}}\:\sum_{k\in\Theta}r_{k}^{BL1}-\kappa\sum_{k\in\Theta}p_{k},
\]
where $r_{k}^{BL1}$ is the instantaneous data rate of user $k$ under
baseline 1; and $\kappa>0$ is used to tradeoff between interruption
probability and average transmit power.

\textbf{Baseline 2 (Q-weighted without RS):} The physical layer is
the same as that in baseline 1. The power control is adaptive to both
CSI and QSI. Specifically, at each time slot, the power control is
obtained by solving the following optimization problem:
\begin{equation}
\underset{\left\{ p_{k}\right\} }{\textrm{max}}\:\sum_{k\in\Theta}\left(W_{H}-Q_{k}\right)^{+}r_{k}^{BL1}-\kappa\sum_{k\in\Theta}p_{k}.\label{eq:QWP}
\end{equation}

\textbf{Baseline 3: (Q-weighted DF):} At each time slot, $2$ users
are selected randomly for transmission from the $4$ users. Then the
full decode and forward (DF) relay scheme proposed in \cite{Simoens_TSP09_MIMOrelay}
is employed at the physical layer to serve the selected $2$ users.
The channel between the BS and RS is modeled by $\mathbf{H}_{BR}=10\mathbf{H}_{W}\in\mathbb{C}^{M\times M}$,
where $\mathbf{H}_{W}$ has i.i.d. complex Gaussian entries of zero
mean and unit variance (i.e., the path gain of the the channel between
the BS and RS is 20dB larger than the user channel $\mathbf{h}_{k}$'s).
In the relay listening phase, the interference among the data streams
of different users is eliminated using ZF transmit beamforming at
the BS and ZF receiving beamforming at the RS. In the BS-RS cooperative
transmission phase, the multi-user interference is eliminated using
joint ZF beamforming at the BS and RS. At each time slot, the power
control is obtained by solving the optimization problem in (\ref{eq:QWP})
with $r_{k}^{BL1}$ replaced by $r_{k}^{BL3}$, the instantaneous
data rate of user $k$ under baseline 3.

\begin{figure}
\begin{centering}
\includegraphics[width=85mm]{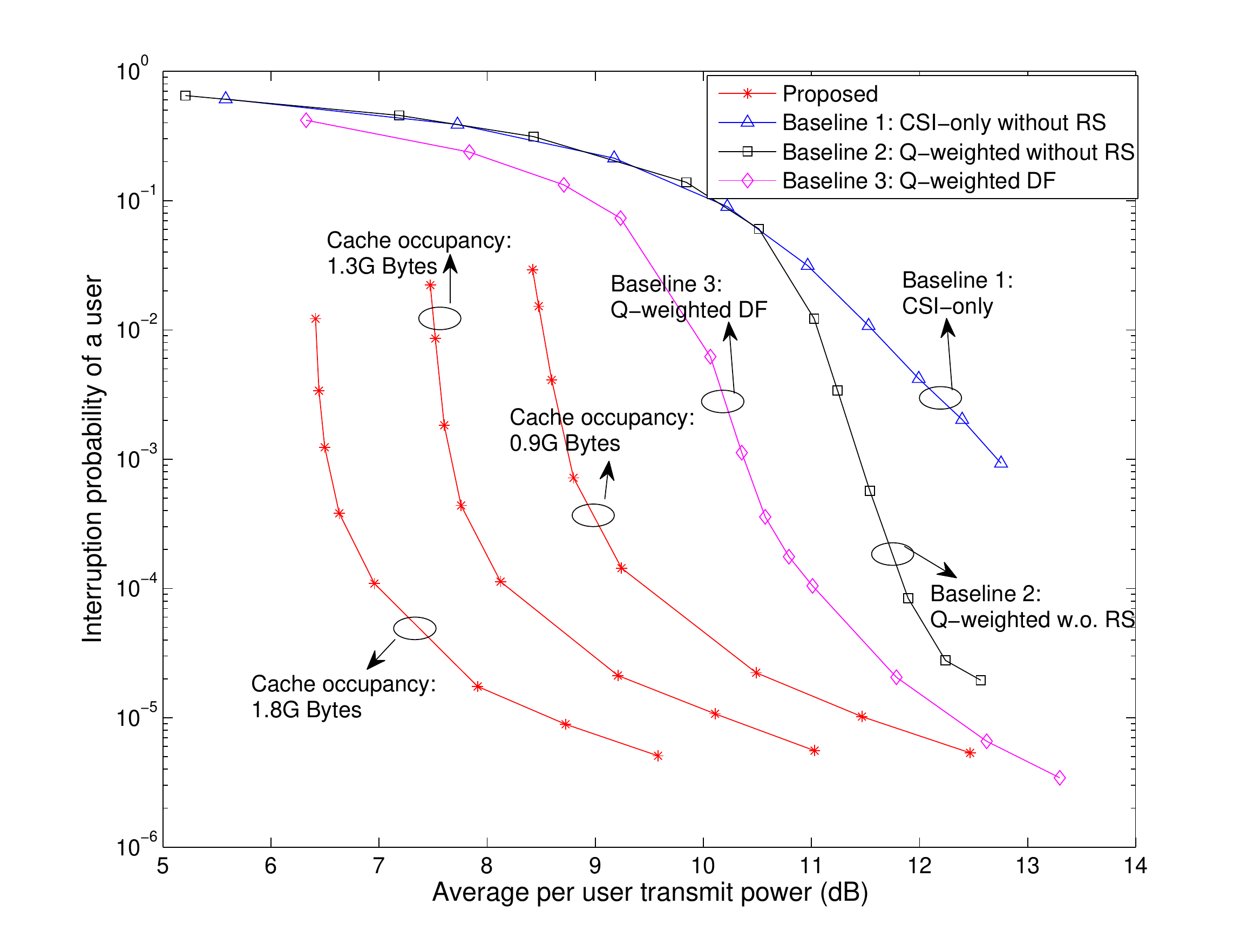}
\par\end{centering}

\caption{\label{fig:ITRprob}Playback interruption probability versus the average
per-user transmit power. }
\end{figure}

\begin{figure}
\begin{centering}
\includegraphics[width=85mm]{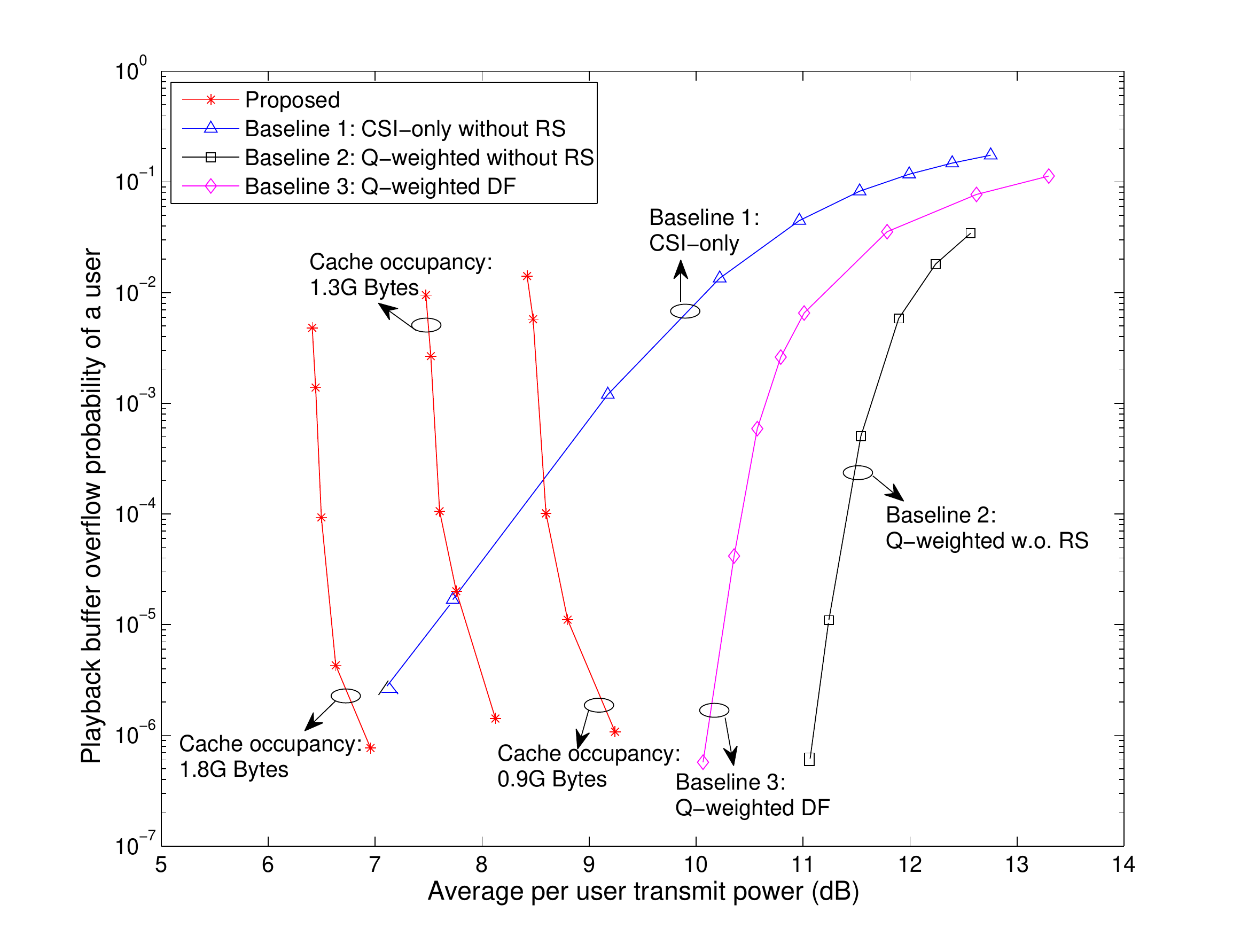}
\par\end{centering}

\caption{\label{fig:PBFprob}Playback overflow probability versus the average
per-user transmit power. }
\end{figure}

\begin{figure}
\begin{centering}
\includegraphics[width=85mm]{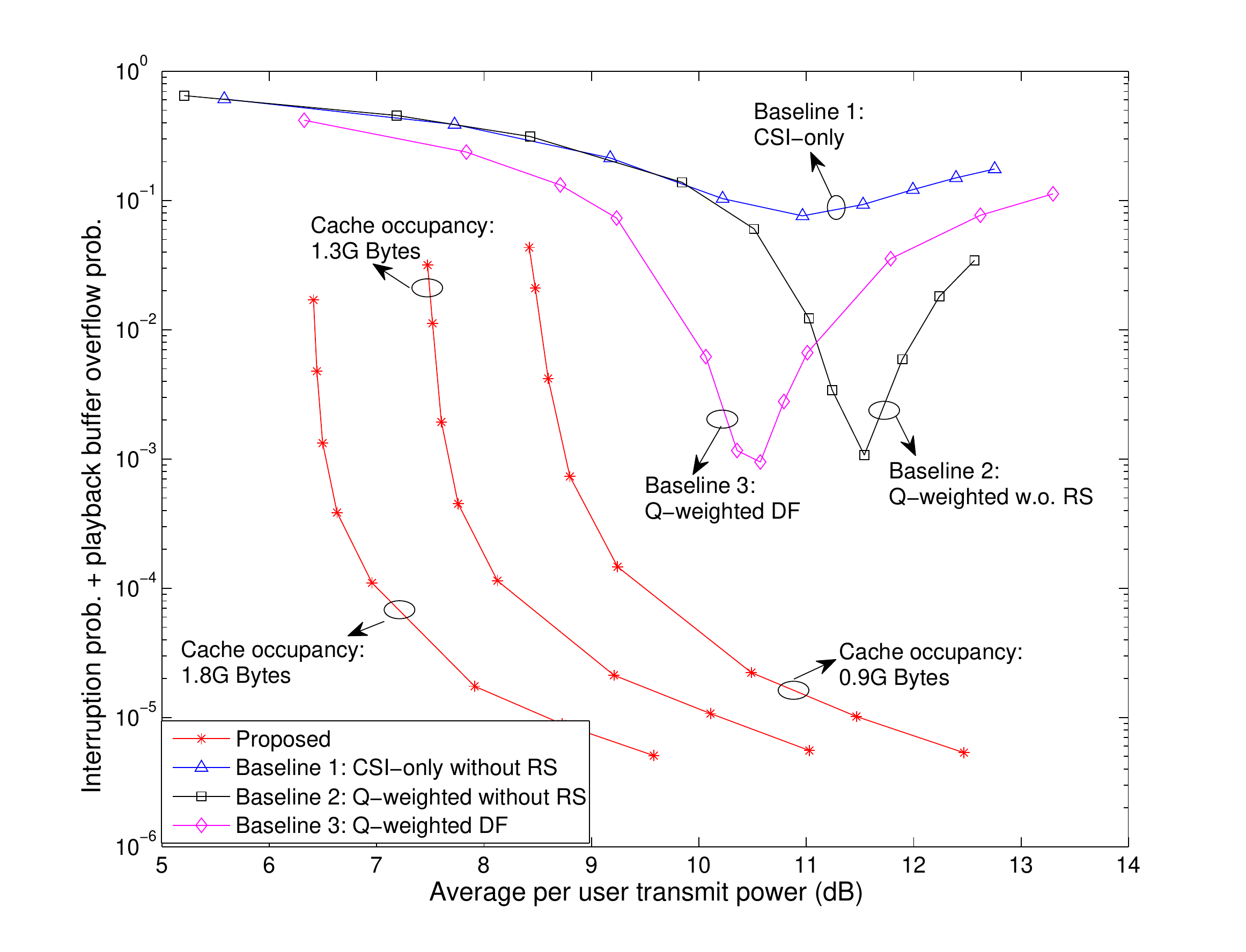}
\par\end{centering}

\caption{\label{fig:IFPPBFprob}Combined interruption probability and playback
overflow probability versus the average per-user transmit power. }
\end{figure}

In the simulations, we let $\beta_{k}=\gamma_{k}=\beta,\forall k$
and vary $\beta,\eta$ to obtain different tradeoff curves. For the
baselines, the tradeoff curves are obtained by varying $\kappa$.
In Fig. \ref{fig:ITRprob} and \ref{fig:PBFprob}, we respectively
plot the playback interruption probability and playback buffer overflow
probability of user $k$ versus the average per-user transmit power%
\footnote{Note that all users have the same interruption / playback buffer probability
and average per-user transmit power since we set $\beta_{k}=\gamma_{k}=\beta,\forall k$
in the simulations.%
} $\textrm{E}\left[p_{k}\right]$. It can be observed that the proposed
solution has a significant performance gain over all baselines, and
the gain increases as more parity bits of the video files are stored
in the RS cache. As the average transmit power increases, both the
interruption probability and the playback buffer overflow probability
of the proposed solution decreases; however, the playback buffer overflow
probability of the baselines increases%
\footnote{This is because the optimization objectives (physical layer throughput)
of the baselines are not properly designed to capture the playback
buffer cost. For example, for baseline 1, the objective function only
contains the sum rate and the average sum transmit power cost. As
$\kappa$ decreases, the average transmit power increases and the
data rate of each user also increases. Since the playback rate is
constant for $Q_{k}\geq W_{L}$, a higher data rate (arrival rate)
leads to a higher playback buffer overflow probability. As a result,
the playback buffer overflow probability increases as the average
transmit power increases ($\kappa$ decreases). Similar observations
can also be made for baseline 2 and baseline 3.%
}. This shows that the system cost function and the associated power
control algorithm must be carefully designed in order to achieve a
good tradeoff between interruption probability, playback buffer overflow
probability and the transmit power cost. For example, in Fig. \ref{fig:ITRprob},
although the interruption probability of the baselines approaches
that of the proposed solution as the average transmit power becomes
large, the corresponding playback buffer overflow probability is also
very large. In Fig. \ref{fig:PBFprob}, although the playback buffer
overflow probability of the baselines is small for small average transmit
power, the corresponding interruption probability is very large. As
a result, the overall combined video streaming performance of the
baselines is much worse than the proposed solution as illustrated
in Fig. \ref{fig:IFPPBFprob}. This demonstrates that the proposed
solution can achieve a much better tradeoff between different system
costs.

\section{Conclusion\label{sec:Conlusion} }

We introduce a cache-enabled opportunistic CoMP framework for wireless
video streaming. By caching a portion of the video files at the relays
(RS), the BS and RSs are able to opportunistically employ CoMP without
expensive backhaul. We first propose a novel MDS-coded random cache
data structure to significantly improve the CoMP opportunities. We
then formulate a two timescale joint optimization problem for power
and cache control. The long-term cache control is used to achieve
the best tradeoff between CoMP opportunities and RS cache size. The
short-term power control is to guarantee the QoS requirements for
given cache control. We derive a closed-form power control solution
and propose a stochastic subgradient algorithm to find the cache control
solution. The proposed solution has low complexity and can achieve
significant performance gain over conventional relay techniques as
demonstrated by numerical simulations.

\appendix

\subsection{Proof of Theorem \ref{thm:Equ_bellman}\label{sub:Proof-of-TheoremEqubellman}}

Following Proposition 4.6.1 of \cite{Bertsekas_MIT07_DynProg}, the
sufficient conditions for optimality of $\mathcal{P}_{I}\left(\mathbf{q},\pi\right)$
is that there exists a $\left(\theta^{*},\left\{ V^{*}\left(\mathbf{Q}\right)\right\} \right)$
that satisfies the following Bellman equation and $V^{*}$ satisfies
the transversality condition in (\ref{eq:transcond}) for all admissible
control policy and initial state $\mathbf{Q}(0)$:
\begin{eqnarray*}
 &  & \theta^{*}\tau+V^{*}\left(\boldsymbol{\chi}\right)\\
 & = & \underset{\left\{ p_{k}\right\} }{\textrm{min}}\left[c\left(\mathbf{Q},\left\{ p_{k}\right\} \right)\tau+\sum_{\boldsymbol{\chi}^{'}}\textrm{Pr}\left[\boldsymbol{\chi}^{'}|\boldsymbol{\chi},\left\{ p_{k}\right\} \right]V^{*}\left(\boldsymbol{\chi}^{'}\right)\right],\\
 & = & \underset{\left\{ p_{k}\right\} }{\textrm{min}}\bigg[c\left(\mathbf{Q},\left\{ p_{k}\right\} \right)\tau+\\
 &  & \sum_{\mathbf{Q}^{'}}\sum_{S^{'},\mathbf{H}^{'}}\textrm{Pr}\left[\mathbf{Q}^{'}|\boldsymbol{\chi},\left\{ p_{k}\right\} \right]\textrm{Pr}\left[S^{'},\mathbf{H}^{'}\right]V^{*}\left(\boldsymbol{\chi}^{'}\right)\bigg].
\end{eqnarray*}
Taking expectation w.r.t. $S,\mathbf{H}$ on both sides of the above
equation and denoting $V^{*}\left(\mathbf{Q}\right)=\textrm{E}\left[\left.V^{*}\left(\boldsymbol{\chi}\right)\right|\mathbf{Q}\right]$,
we obtain the equivalent Bellman equation in (\ref{eq:Equ_Bellman})
in Theorem \ref{thm:Equ_bellman}.

\subsection{Proof of Corollary \ref{cor:Approximate-Bellman-EquationFor}\label{sub:Proof-of-CorollaryApproxACOE}}

Let $\mathbf{Q}^{'}=\left(Q_{1}^{'},...,Q_{2M}^{'}\right)=\mathbf{Q}\left(t+1\right)$
and $\mathbf{Q}=\mathbf{Q}\left(t\right)$. According to the dynamics
of the playback buffer in (\ref{eq:Qdyn}), we have $Q_{k}^{'}=Q_{k}+\left(r_{k}\left(g_{k},p_{k}\right)-\mu_{k}\left(Q_{k}\right)\right)\tau$.
If $V(\mathbf{Q})\in\mathcal{C}^{1}\left(\mathbb{R}_{+}^{2M}\right)$,
we have the following first order Taylor expansion on $V\left(\mathbf{Q}^{'}\right)$
in (\ref{eq:Equ_Bellman}):
\begin{eqnarray*}
 &  & V\left(\mathbf{Q}^{'}\right)=V\left(\mathbf{Q}\right)+\nu G_{\boldsymbol{\chi}}\left(V,\left\{ p_{k}\right\} \right)\\
 &  & +\sum_{k=1}^{2M}\frac{\partial V\left(\mathbf{Q}\right)}{\partial Q_{k}}\left(r_{k}\left(\mathbf{g}_{k},p_{k}\right)-\mu_{k}\left(Q_{k}\right)\right)\tau.
\end{eqnarray*}
where $G_{\boldsymbol{\chi}}$ is some smooth function and $\nu=o\left(1\right)$.
For notation convenience, let $J_{\boldsymbol{\chi}}\left(\theta,V,\left\{ p_{k}\right\} \right)$
denote the \textit{Bellman operator}:
\begin{eqnarray*}
J_{\boldsymbol{\chi}}\left(\theta,V,\left\{ p_{k}\right\} \right) & = & c\left(\mathbf{Q},\left\{ p_{k}\right\} \right)+\nu G_{\boldsymbol{\chi}}\left(V,\left\{ p_{k}\right\} \right)-\theta\\
 & + & \sum_{k=1}^{2M}\frac{\partial V\left(\mathbf{Q}\right)}{\partial Q_{k}}\left(r_{k}\left(\mathbf{g}_{k},p_{k}\right)-\mu_{k}\left(Q_{k}\right)\right).
\end{eqnarray*}
Denote
\[
J_{\mathbf{Q}}^{*}\left(\theta,V\right)=\textrm{E}\left[\left.\underset{\left\{ p_{k}\right\} }{\textrm{min}}J_{\boldsymbol{\chi}}\left(\theta,V,\left\{ p_{k}\right\} \right)\right|\mathbf{Q}\right].
\]
By definition, we have
\[
J_{\mathbf{Q}}^{*}\left(\theta^{*},V^{*}\right)=0,\:\forall\mathbf{Q}\in\boldsymbol{\mathcal{Q}}.
\]

Similarly, define \textit{approximate Bellman operator}:
\begin{eqnarray*}
\hat{J}_{\boldsymbol{\chi}}\left(\theta,V,\left\{ p_{k}\right\} \right) & = & c\left(\mathbf{Q},\left\{ p_{k}\right\} \right)-\theta\\
 & + & \sum_{k=1}^{2M}\frac{\partial V\left(\mathbf{Q}\right)}{\partial Q_{k}}\left(r_{k}\left(\mathbf{g}_{k},p_{k}\right)-\mu_{k}\left(Q_{k}\right)\right).
\end{eqnarray*}
If $\left(\theta,V\right)$ satisfies the approximate Bellman equation
in (\ref{eq:approxACOE}), we have
\begin{equation}
\hat{J}_{\mathbf{Q}}^{*}\left(\theta,V\right)=0,\:\forall\mathbf{Q}\in\boldsymbol{\mathcal{Q}},\label{eq:JhadQ0}
\end{equation}
where $\hat{J}_{\mathbf{Q}}^{*}\left(\theta,V\right)=\textrm{E}\left[\left.\underset{\left\{ p_{k}\right\} }{\textrm{min}}\hat{J}_{\boldsymbol{\chi}}\left(\theta,V,\left\{ p_{k}\right\} \right)\right|\mathbf{Q}\right]$.
The proof relies on the following Lemma.
\begin{lem}
\label{lem:J12}If $\left(\theta,V\right)$ satisfies the approximate
Bellman equation, i.e., $\hat{J}_{\mathbf{Q}}^{*}\left(\theta,V\right)=0$,
then $J_{\mathbf{Q}}^{*}\left(\theta,V\right)=o\left(1\right),\forall\mathbf{Q}\in\boldsymbol{\mathcal{Q}}$. 
\end{lem}

Lemma \ref{lem:J12} follows straightforward from the definitions
of $J_{\mathbf{Q}}^{*}$, $\hat{J}_{\mathbf{Q}}^{*}$ and the proof
is omitted for conciseness.

Finally, we use contradiction to show that $\theta=\theta^{*}+o\left(1\right)$
and $V\left(\mathbf{Q}\right)=V^{*}\left(\mathbf{Q}\right)+o\left(1\right)$
if $\left(\theta,V\right)$ satisfies the approximate Bellman equation.
Suppose for some $\mathbf{Q}^{'}$, we have $V\left(\mathbf{Q}^{'}\right)=V^{*}\left(\mathbf{Q}^{'}\right)+\mathcal{O}\left(1\right)$.
Since $J_{\mathbf{Q}}^{*}\left(\theta,V\right)=o\left(1\right),\forall\mathbf{Q}\in\boldsymbol{\mathcal{Q}}$,
as $\tau\rightarrow0$, we have $J_{\mathbf{Q}}^{*}\left(\theta,V\right)=0,\forall\mathbf{Q}\in\boldsymbol{\mathcal{Q}}$
and $\left(\theta,V\right)$ satisfy the transversality condition
in (\ref{eq:transcond}). However, $V\left(\mathbf{Q}^{'}\right)\neq V^{*}\left(\mathbf{Q}^{'}\right)$
due to the assumption that $V\left(\mathbf{Q}^{'}\right)=V^{*}\left(\mathbf{Q}^{'}\right)+\mathcal{O}\left(1\right)$.
This contradicts with the condition that $\left(\theta^{*},V^{*}\right)$
is the unique solution of $J_{\mathbf{Q}}^{*}\left(\theta^{*},V^{*}\right)=0,\forall\mathbf{Q}\in\boldsymbol{\mathcal{Q}}$
and the transversality condition in (\ref{eq:transcond}). Hence,
we must have $V\left(\mathbf{Q}\right)=V^{*}\left(\mathbf{Q}\right)+o\left(1\right),\forall\mathbf{Q}\in\boldsymbol{\mathcal{Q}}$.
Similarly, we can establish $\theta=\theta^{*}+o\left(1\right)$.

\subsection{Proof of Lemma \ref{lem:PDF-of-EFC} \label{Proof-of-Lemma-PDFCH}}

Conditioned on $S=1$, we have $g_{k}=\left|\mathbf{h}_{k}^{\dagger}\mathbf{v}_{k}\right|^{2}$.
Using similar proof as that of \cite{Li_TIT06_MMSEsinrdist}, it can
be shown that $\left|\mathbf{h}_{k}^{\dagger}\mathbf{v}_{k}\right|^{2}$
is a Gamma random variable, $\left|\mathbf{h}_{k}^{\dagger}\mathbf{v}_{k}\right|^{2}\sim G\left(1,1\right)$
as elaborated below. Let $\mathbf{H}_{\left(-k\right)}=\left[\mathbf{h}_{k^{'}}\right]_{k^{'}\in\Theta\backslash\left\{ k\right\} }$
and consider the singular value decomposition (SVD): $\mathbf{H}_{\left(-k\right)}=\mathbf{U}\mathbf{D}\mathbf{V}^{\dagger}$,
where $\mathbf{U}\in\mathbb{C}^{\left(2M-1\right)\times\left(2M-1\right)}$
is unitary, $\mathbf{D}\in\mathbb{R}^{\left(2M-1\right)\times\left(2M-1\right)}$,
$\mathbf{V}\in\mathbb{C}^{2M\times\left(2M-1\right)}$ is semi-unitary.
Let $\mathbf{v}_{c}\in\mathbb{C}^{2M\times1}$ be the orthogonal complement
of $\mathbf{V}$, i.e., $\mathbf{v}_{c}^{\dagger}\mathbf{V}=\mathbf{0}$
and $\left\Vert \mathbf{v}_{c}\right\Vert =1$. Then according to
the definition of ZF beamforming vector, we have $\mathbf{v}_{k}=\mathbf{v}_{c}\mathbf{v}_{c}^{\dagger}\mathbf{h}_{k}/\left|\mathbf{v}_{c}^{\dagger}\mathbf{h}_{k}\right|$.
Hence, $g_{k}=\left|\mathbf{h}_{k}^{\dagger}\mathbf{v}_{c}\mathbf{v}_{c}^{\dagger}\mathbf{h}_{k}/\left|\mathbf{v}_{c}^{\dagger}\mathbf{h}_{k}\right|\right|^{2}=\left|\mathbf{v}_{c}^{\dagger}\mathbf{h}_{k}\right|^{2}$.
Note that conditioned on $\mathbf{H}_{\left(-k\right)}$, $\mathbf{v}_{c}^{\dagger}\mathbf{h}_{k}\sim\mathcal{CN}\left(0,1\right)$.
Since the conditional distribution of $\mathbf{v}_{c}^{\dagger}\mathbf{h}_{k}$
given $\mathbf{H}_{\left(-k\right)}$ is independent of $\mathbf{H}_{\left(-k\right)}$,
$\mathcal{CN}\left(0,1\right)$ is also the unconditional distribution
of $\mathbf{v}_{c}^{\dagger}\mathbf{h}_{k}$. Therefore, conditioned
on $S=1$, $g_{k}=\left|\mathbf{v}_{c}^{\dagger}\mathbf{h}_{k}\right|^{2}\sim G\left(1,1\right)$
and the PDF is $h_{g_{k}|S=1}\left(x\right)=e^{-x}$. Similarly, it
can be shown that conditioned on $S=0,k\in\Theta$, the PDF of $g_{k}$
is $h_{g_{k}|S=0,k\in\Theta}\left(x\right)=e^{-x}$; and conditioned
on $S=0,k\notin\Theta$, the PDF of $g_{k}$ is $h_{g_{k}|S=0,k\notin\Theta}\left(x\right)=\delta\left(x\right)$.
Therefore the unconditional PDF of $g_{k}$ is $h_{g_{k}}\left(x\right)=\Pr\left(k\in\Theta\right)e^{-x}+\Pr\left(k\notin\Theta\right)\delta\left(x\right)=\left(\frac{1+q_{\textrm{min}}}{2}\right)e^{-x}+\left(\frac{1-q_{\textrm{min}}}{2}\right)\delta\left(x\right).$

\subsection{Proof of Theorem \ref{thm:Per-user-Bellman-Solution}\label{sub:Proof-of-TheoremCFbellman}}

Let $g\left(x,f_{k}\right)$ denote the L.H.S. of (\ref{eq:ODEVQ}).
It can be shown that for fixed $x\in\left[0,Q^{\circ}\right]$, $g\left(x,f_{k}\right)$
is strictly increasing w.r.t. $f_{k}$ over $f_{k}\in\left(-\infty,\tilde{\lambda}\left(x\right)\right]$
and $g\left(x,\tilde{\lambda}\left(x\right)\right)\geq\tilde{\theta}_{k}$.
Then it follows that $g\left(x,f_{k}\right)=\tilde{\theta}_{k}$ has
a unique solution over $f_{k}\in\left(-\infty,\tilde{\lambda}\left(x\right)\right]$.
Similarly, it can be shown that for fixed $x\in\left(Q^{\circ},\infty\right)$,
$g\left(x,f_{k}\right)$ is strictly decreasing w.r.t. $f_{k}$ over
$f_{k}\in\left[\tilde{\lambda}\left(x\right),\infty\right]$ and $g\left(x,\tilde{\lambda}\left(x\right)\right)\geq\tilde{\theta}_{k}$.
Then it follows that $g\left(x,f_{k}\right)=\tilde{\theta}_{k}$ has
a unique solution over $f_{k}\in\left[\tilde{\lambda}\left(x\right),\infty\right]$.
This completes the proof for Result 1) and 2).

The following Lemma is required to prove Result 3). 
\begin{lem}
[Decomposed Bellman Equation]\label{lem:Decomposed-Bellman-Equation}Suppose
that for any $1\leq k\leq2M$, there exist $\theta_{k}$ and $V_{k}\left(Q_{k}\right)\in C^{1}\left(\mathbb{R}_{+}\right)$
that solve the following per-user approximate Bellman equation:
\begin{eqnarray}
\theta_{k} & = & \textrm{E}\bigg[\underset{p_{k}}{\textrm{min}}\big[\hat{c}_{k}\left(Q_{k}\right)+p_{k}\nonumber \\
 &  & +\left.V_{k}^{'}\left(Q_{k}\right)\left(r_{k}\left(g_{k},p_{k}\right)-\mu_{k}\left(Q_{k}\right)\right)\big]\right|Q_{k}\bigg]\label{eq:per-user-approxACOE}
\end{eqnarray}
for all $Q_{k}\in\mathcal{Q}$, where $V_{k}^{'}\left(Q_{k}\right)=\frac{\partial V_{k}\left(Q_{k}\right)}{\partial Q_{k}}$.
Furthermore, for all admissible control policy $\Omega$ and initial
queue state $\mathbf{Q}(0)$, the transversality condition in (\ref{eq:transcond})
is satisfied for $V\left(\mathbf{Q}\right)\triangleq\sum_{k=1}^{2M}V_{k}\left(Q_{k}\right),\:\mathbf{Q}\in\boldsymbol{\mathcal{Q}}$.
Let $\theta\triangleq\sum_{k=1}^{2M}\theta_{k}$. Then $\left(\theta,\left\{ V\left(\mathbf{Q}\right)\right\} \right)$
satisfy the approximate Bellman equation in Corollary \ref{cor:Approximate-Bellman-EquationFor}.
\end{lem}

Lemma \ref{lem:Decomposed-Bellman-Equation} can be proved using the
fact that the dynamics of the playback buffer at the $2M$ users are
decoupled. The details are omitted for conciseness.

The optimal power control policy that attains the minimum in (\ref{eq:per-user-approxACOE})
is given by (\ref{eq:optpow}) with $\tilde{f}_{k}\left(x\right)$
replaced by $V_{k}^{'}\left(Q\right)$. Substituting (\ref{eq:optpow})
into (\ref{eq:per-user-approxACOE}) and calculating the expectations
using Lemma \ref{lem:PDF-of-EFC}, (\ref{eq:per-user-approxACOE})
can be transformed into (\ref{eq:ODEVQ}) with $x$ replaced by $Q_{k}$,
$f_{k}$ replaced by $V_{k}^{'}\left(Q\right)$ and $\tilde{\theta}_{k}$
replaced by $\theta_{k}$. Using the above fact, it can be verified
that $\tilde{\theta}_{k}$ and $\tilde{V}_{k}\left(Q_{k}\right)\triangleq\int_{Q_{k}^{\circ}}^{Q_{k}}\tilde{f}_{k}\left(x\right)dx,\:\forall Q_{k}\in\mathcal{Q}$
satisfy the per-user approximate Bellman equation (\ref{eq:per-user-approxACOE}).
Moreover, it can be verified that there exists $Q_{U}>\underset{k}{\textrm{max}}\: Q_{k}^{\circ}$
and $\epsilon>0$, such that $\textrm{E}\left[\left.r_{k}\left(g_{k},p_{k}\right)\right|Q_{k}>Q_{U}\right]\leq\mu_{0}-\epsilon,\forall k$
under the power control policy in (\ref{eq:optpow}). This implies
that the control policy (\ref{eq:optpow}) is admissible. Finally,
it can be shown that there exists $C>0$ such that $\left|\tilde{f}_{k}\left(x\right)\right|\leq C,\forall x\in\mathcal{Q}$,
i.e., $\left|\tilde{f}_{k}\left(x\right)\right|$ is bounded for all
$x\in\mathcal{Q}$. Hence, $\left|\tilde{V}_{k}\left(Q_{k}\right)\right|\leq\left|\int_{Q_{k}^{\circ}}^{Q_{k}}\left|\tilde{f}_{k}\left(x\right)\right|dx\right|\leq C\left|Q_{k}-Q_{0}\right|$,
which implies that $\tilde{V}\left(\mathbf{Q}\right)\triangleq\sum_{k=1}^{2M}\tilde{V}_{k}\left(Q_{k}\right)$
satisfies the transversality condition in (\ref{eq:transcond}). Then
it follows from Lemma \ref{lem:Decomposed-Bellman-Equation} that
$\theta^{*}=\tilde{\theta}+o\left(1\right)$ and $V^{*}\left(\mathbf{Q}\right)=\tilde{V}\left(\mathbf{Q}\right)+o\left(1\right),\:\forall\mathbf{Q}\in\boldsymbol{\mathscr{Q}}$.

\subsection{Proof of Proposition \ref{prop:cacheunderflow}\label{sub:Proof-of-PropositionUF}}

Let $\Omega$ denote the power control policy in Table \ref{tab:Algpow}.
Let $\overline{r}_{k}^{A}=\textrm{E}\left[\left.r_{k}\left(\mathbf{g}_{k},\Omega_{k}\left(\mathbf{Q},S,\mathbf{H}\right)\right)\right|S=1\right]$
and $\overline{r}_{k}^{B}=\textrm{E}\left[\left.r_{k}\left(\mathbf{g}_{k},\Omega_{k}\left(\mathbf{Q},S,\mathbf{H}\right)\right)\right|k\in\Theta\right]$.
It can be verified that $\overline{r}_{k}^{A}=\overline{r}_{k}^{B}$.
Let $T_{S}$ denote the total number of time slots used to transmit
current segment for user $k$. Since $\Omega$ is a unichain policy
and the queueing system under $\Omega$ is stable, the induced Markov
chain $\boldsymbol{\chi}\left(t\right)$ has a single recurrent class.
Moreover, $\boldsymbol{\chi}\left(t\right)$ is ergodic over its recurrent
class \cite{Bertsekas_MIT07_DynProg}. Hence, we have 
\begin{equation}
\overline{r}_{k}^{B}=\underset{T_{S}\rightarrow\infty}{\textrm{lim}}\frac{L_{S}/T_{S}}{\frac{1}{T_{S}}\sum_{t=1}^{T_{S}}1\left(k\in\Theta\left(t\right)\right)}=\frac{2\mu_{0}\tau}{1+q_{\textrm{min}}}.\label{eq:rkB}
\end{equation}
The cache underflow occurs if the total number of the transmitted
cached parity bits is more than $\frac{2q_{\pi_{k}}L_{S}}{\left(1+q_{\pi_{k}}\right)}$.
In this case, there exists $\delta>0$ such that during the first
$T_{S}\left(1-\delta\right)$ time slots, the number of the transmitted
cached parity bits is equal to $\frac{2q_{\pi_{k}}L_{S}}{\left(1+q_{\pi_{k}}\right)}$.
In the following, we use contradiction to show that $\delta\rightarrow0$
as $T_{S}\rightarrow\infty$ (or equivalently, $L_{S}\rightarrow\infty$),
which implies that the cache underflow probability tends to zero as
$L_{S}\rightarrow\infty$. Suppose that $\delta>0$ as $T_{S}\rightarrow\infty$.
Then following similar analysis for (\ref{eq:rkB}), we have
\begin{eqnarray*}
\overline{r}_{k}^{A} & = & \underset{T_{S}\rightarrow\infty}{\textrm{lim}}\frac{2q_{\pi_{k}}L_{S}/T_{S}}{\left(1+q_{\pi_{k}}\right)\frac{1}{T_{S}}\sum_{t=1}^{T_{S}\left(1-\delta\right)}1\left(S\left(t\right)=1\right)}\\
 & = & \frac{2q_{\pi_{k}}\mu_{0}\tau}{\left(1+q_{\pi_{k}}\right)\left(1-\delta\right)q_{\textrm{min}}}>\overline{r}_{k}^{B},
\end{eqnarray*}
which contradicts with $\overline{r}_{k}^{A}=\overline{r}_{k}^{B}$.
Hence, we must have $\delta\rightarrow0$ as $T_{S}\rightarrow\infty$.

\subsection{Proof of Theorem \ref{thm:Asymptotic-Approximation-of-Cbar}\label{Proof-of-TheoremCbar}}

We first prove the following Lemma.
\begin{lem}
\label{lem:forCbar}As $\overline{\mu}_{0}\triangleq\frac{\mu_{0}}{B_{W}}\rightarrow\infty$,
we have $\tilde{\theta}=\hat{C}\left(\mathbf{q},\pi\right)+\mathcal{O}(1)$
and $\hat{C}\left(\mathbf{q},\pi\right)=\mathcal{O}\left(e^{\overline{\mu}_{0}}\right)$,
where $\tilde{\theta}$ is given in Theorem \ref{thm:Per-user-Bellman-Solution}.
\end{lem}

\begin{IEEEproof}
For convenience, let $\overline{\lambda}_{k}=-\frac{B_{W}\tilde{\lambda}_{k}\left(Q_{k}^{\circ}\right)}{\textrm{ln}2}$,
where $\tilde{\lambda}_{k}\left(Q_{k}^{\circ}\right)$ is the unique
solution of (\ref{eq:Lamtutadef}) for $x=Q_{k}^{\circ}$. Then $\overline{\lambda}_{k}$
is the unique solution of 
\begin{equation}
\phi\left(\lambda\right)\triangleq\frac{\xi}{\textrm{ln}2}E_{1}\left(\frac{1}{\lambda}\right)-\overline{\mu}_{0}=0,\label{eq:Lamtuta}
\end{equation}
 w.r.t. $\lambda\in\mathbb{R}_{+}$. Let $\hat{\lambda}_{k}=e^{-a_{1}+\frac{\overline{\mu}_{0}\textrm{ln}2}{\xi}}$
denote the unique solution of
\[
\hat{\phi}\left(\lambda\right)\triangleq\frac{\xi}{\textrm{ln}2}\left(\textrm{log}\left(\lambda\right)+a_{1}\right)-\overline{\mu}_{0}=0,
\]
w.r.t. $\lambda\in\mathbb{R}_{+}$. It can be verified that $\overline{\lambda}_{k}=\mathcal{O}\left(e^{\overline{\mu}_{0}}\right)$
and $\hat{\lambda}_{k}=\mathcal{O}\left(e^{\overline{\mu}_{0}}\right)$,
from which it can be shown that $\left|\phi\left(\lambda\right)-\hat{\phi}\left(\lambda\right)\right|=\mathcal{O}\left(\overline{\mu}_{0}e^{-\overline{\mu}_{0}}\right),\forall\lambda\in\left\{ \overline{\lambda}_{k},\hat{\lambda}_{k}\right\} $.
Then it follows that 
\begin{equation}
\left|\phi\left(\overline{\lambda}_{k}\right)-\phi\left(\hat{\lambda}_{k}\right)\right|=\mathcal{O}\left(\overline{\mu}_{0}e^{-\overline{\mu}_{0}}\right).\label{eq:dltfai}
\end{equation}
Combining (\ref{eq:dltfai}) and the fact that $\phi\left(\lambda\right)$
is a strictly increasing and differentiable function of $\lambda$
over $\mathbb{R}_{+}$ with $\phi^{'}\left(\lambda\right)=\mathcal{O}\left(e^{-\overline{\mu}_{0}}\right)$
for $\lambda=\mathcal{O}\left(e^{\overline{\mu}_{0}}\right)$, we
have 
\begin{equation}
\left|\overline{\lambda}_{k}-\hat{\lambda}_{k}\right|=\mathcal{O}\left(\overline{\mu}_{0}\right).\label{eq:dltlam}
\end{equation}
Using $\overline{\lambda}_{k}=\mathcal{O}\left(e^{\overline{\mu}_{0}}\right)$,
(\ref{eq:dltlam}) and the definition of $\tilde{\theta}_{k}$ in
(\ref{eq:Thetatuta}), it can be shown that
\[
\tilde{\theta}_{k}=\xi e^{-a_{1}+\frac{\mu_{0}\textrm{ln}2}{B_{W}\xi}}-(a_{2}+e)\xi+e+\hat{c}_{k}\left(Q_{k}^{\circ}\right)+\mathcal{O}\left(\overline{\mu}_{0}\right),
\]
from which it follows that $\tilde{\theta}=\hat{C}\left(\mathbf{q},\pi\right)+\mathcal{O}(\overline{\mu}_{0})$.
The result that $\hat{C}\left(\mathbf{q},\pi\right)=\mathcal{O}\left(e^{\overline{\mu}_{0}}\right)$
follows directly from the definition of $\hat{C}\left(\mathbf{q},\pi\right)$.
\end{IEEEproof}

Then Theorem \ref{thm:Asymptotic-Approximation-of-Cbar} follows directly
from Theorem \ref{thm:Per-user-Bellman-Solution} and Lemma \ref{lem:forCbar}.


\end{document}